\documentclass[aps,prb,twocolumn,floats]{revtex4}

\usepackage{amssymb}
\usepackage{amsmath}
\usepackage{graphicx}
\usepackage{bm}
\usepackage{mdframed}
\usepackage{color}
\usepackage{gensymb}
\usepackage{comment}

\begin{document}

\bibliographystyle{prsty}

\title{Transport theory and spin-transfer physics for frustrated magnets}

\author{Ricardo Zarzuela$^{1}$ and Jairo Sinova$^{1,2}$}

\affiliation{$^{1}$Institut f\"{u}r Physik, Johannes Gutenberg Universit\"{a}t Mainz, D-55099 Mainz, Germany \\$^{2}$Institute of Physics Academy of Sciences of the Czech Republic, Cukrovarnick\'{a} 10, 162 00 Praha 6, Czech Republic
}

\begin{abstract}
We study the electron dynamics in magnetic conductors with frustrated interactions dominated by isotropic exchange. We present a transport theory for itinerant carriers built upon the (single-band) doped Hubbard model and the slave-boson formalism, which incorporates the spin-exchange with the magnetically frustrated background into the representation of electron operators in a clear and controllable way. We also formulate hydrodynamic equations for the itinerant charge and spin degrees of freedom, whose currents contain new contributions that depend on the spatiotemporal variations of the order parameter of the frustrated magnet, which are described by Yang-Mills fields. Furthermore, we elucidate the transfer of angular momentum from the itinerant charge fluid to the magnet (i.e., the spin-transfer torque) via reciprocity arguments. A detailed microscopic derivation of our effective theory is also provided for one of the simplest models of frustrated magnetism, namely the Heisenberg antiferromagnet on a triangular lattice. Our findings point towards the possibility of previously unanticipated Hall physics in these frustrated platforms.
\end{abstract}
\maketitle

\section{Introduction}

Magnetic systems with frustrated interactions have come to the forefront of Condensed Matter Physics due, in part, to the emergence of phases exhibiting a highly degenerate ground-state manifold and the absence of collinear order. Their unconventional spin excitations are gathering a lot of attention recently, especially in the context of (topological) transport physics \cite{Yamashita-NatPhys2009,Balents-Nature2010,Han-Nature2012,Nasu-PRL2017,Udagawa-PRL2019,Minakawa-PRL2020}. An important (universal) class of frustrated systems are those whose magnetic interactions are dominated by isotropic exchange, where spin glasses\cite{Binder-RMP1986}, amorphous magnets in the spin-correlated phase\cite{Chudnovsky-PRB1982,Chudnovsky-PRB1986} and multilattice antiferromagnets\cite{Dombre-PRB1989} belong to. These systems represent, in their featureless version, one of the simplest realizations of a noncollinear magnet, and at macroscopic scales are described by a SO(3)-order parameter emerging from spin-spin correlations between two possible (ground-state) spin configurations\cite{Henley-AnnPhys1984}. Furthermore, because the long-wavelength spin excitations in this class of materials are generically described by the O(4) nonlinear $\sigma$ model\cite{Azaria-PRL1992,Chubukov-PRL1994}, their spin dynamics are akin to those in antiferromagnets\cite{Footnote1} except for the absence of (geometrical) constraints between the macroscopic magnetization and the order parameter.

Advances in spintronics facilitate the efficient manipulation of local magnetic moments, offering suitable electric- and thermal-based probes to explore the rich interplay among spin, electronic and phononic degrees of freedom present in the aforementioned platforms. For instance, striking transport phenomena has been recently observed in some families of intercalated transition-metal dichalcogenides (epitomized by the Fe$_{1/3}$NbS$_{2}$ compound)\cite{Doyle-2019,Nair-NatMat2020,Little-NatMat2020}, which are contributing to the current revival of interest in glassy antiferromagnets. Also of interest are recent proposals for probing the superfluid spin transport and thermal diffusion of Shankar skyrmions in magnetically frustrated insulators (in the exchange-dominated limit) based on conventional magnetotransport measurements \cite{Ochoa-PRB2018,Zarzuela-PRB2019}. Even though enormous theoretical effort has been made to understand transport phenomena in the insulating scenario, no satisfactory progress has been achieved so far in the conducting counterpart. In this regard, the relatively unexplored family of conducting magnets with frustrated (exchange-dominated) interactions constitutes an outstanding playground to exploit and push forward the field of topological spintronics, making the development of a unified theory for charge and spin transport in these systems imperative.

In this work we construct an effective low-energy long-wavelength theory for itinerant carriers flowing within a magnetically frustrated conductor, based on the doped Hubbard model. We utilize the slave-boson formalism, widely applied in the field of strongly correlated systems, to represent the electron operators. This formalism naturally accounts for the spin-exchange with the localized (spin) degrees of freedom (those described in Sec.~\ref{Frust_mag}), by giving rise to a curved spin-space geometry for the itinerant carriers. One of our main findings is a term in the effective Hamiltonian coupling the spin current of itinerant carriers to the magnetization currents stemming from the topologically nontrivial magnetic background, see Sec.~\ref{Dyn_itinerant_car}. By exploiting the hydrodynamic features of nonrelativistic Yang-Mills theories \cite{Jin-JPA2006}, we show in Sec.~\ref{Hydro} that this coupling term, in particular, yields an extra contribution to both itinerant charge and spin currents, which originates in the noncoplanar nature of the low-lying spin configurations intrinsic to frustrated magnets. As we elucidate in Sec.~\ref{STT}, these contributions, in the adiabatic limit for spin dynamics, determine the transfer of angular momentum from the charge fluid to the magnetic system (i.e, the spin-transfer torque). In Sec.~\ref{ref:Sec2DHA} we provide a microscopic derivation of the aforementioned effective theory for one of the simplest models of frustrated magnetism, namely the triangular Heisenberg antiferromagnet. Sec.~\ref{Disc} is devoted to discussing our results, conclusions and prospective work, including the possibility of Hall physics in these systems. We complement our work with an Appendix describing the principles of the slave-boson method. Hereafter, bold denotes vectors belonging to the spin space and $\vec{\,}\,$ denotes vectors in the real space. Furthermore, $\cdot$ and $\circ$ will denote the dot product in real and spin space, respectively.

\section{Symmetries and the order-parameter manifold of frustrated magnets}
\label{Frust_mag}

\begin{figure}
\begin{center}
\includegraphics[width=1.0\columnwidth]{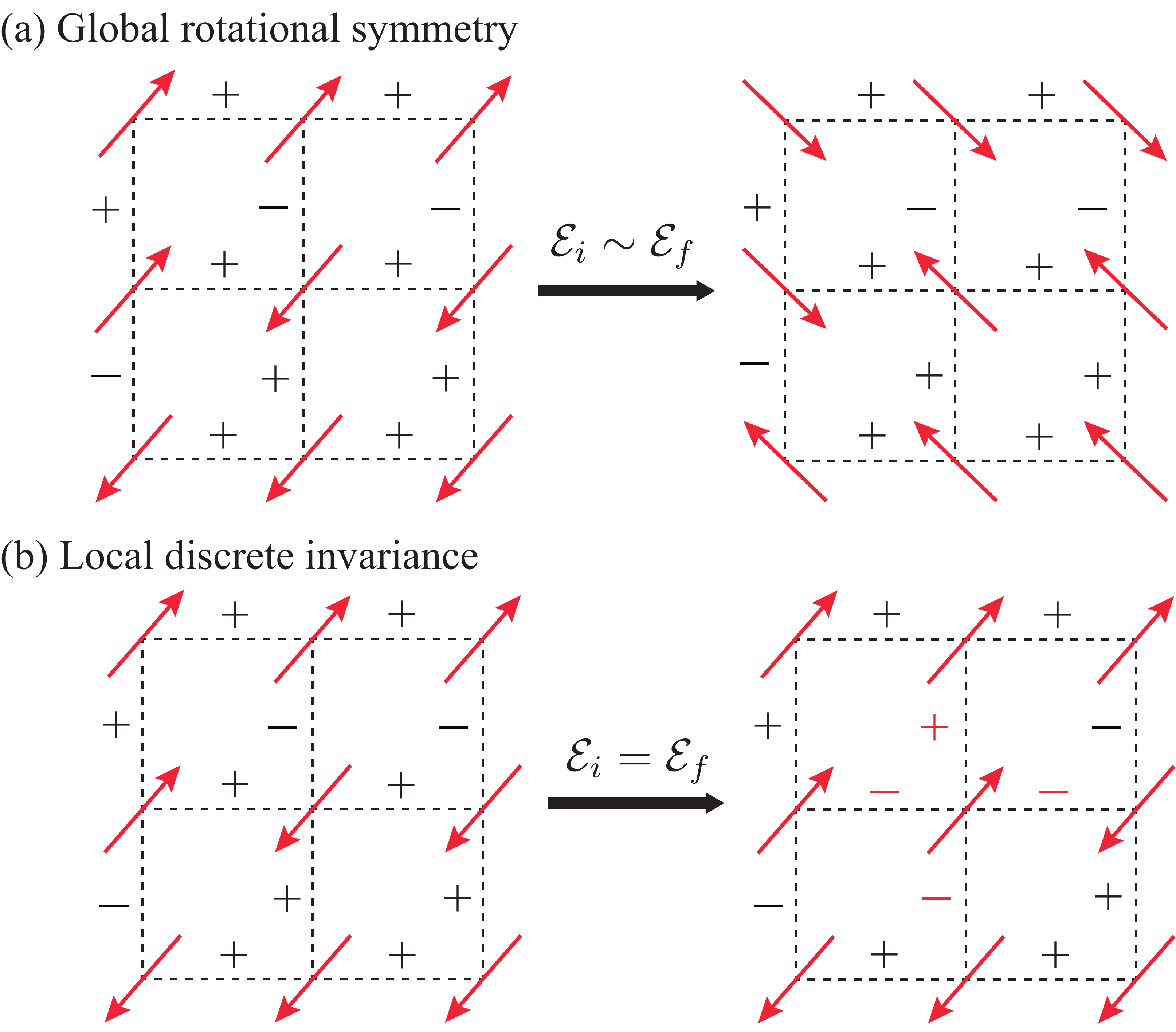}
\caption{Main symmetries of frustrated magnets, namely (a) global rotational symmetry and (b) local discrete invariance. Red arrows represent the lattice spins and '$+/-$' denotes the sign of the exchange coupling constant on each lattice bond. $\mathcal{E}_{i(f)}$ stands for the energy of the initial (final) spin configuration. In the case of global rotational symmetry, (relativistic) magnetocrystalline anisotropies lift weakly the degeneracy between the initial and final states. The choice of a square lattice is merely illustrative.}
\vspace{-0.5cm} 
\label{Fig1}
\end{center}
\end{figure}

Magnetic frustration is rooted in the absence of spin configurations satisfying simultaneously all bond interactions of the lattice. Spin glasses, amorphous magnets and multilattice antiferromagnets represent remarkable instances of magnetic systems with frustrated interactions dominated by isotropic exchange, referred to as \textit{frustrated magnets} hereafter. This class of materials generically exhibits global rotational symmetry (in the spin space) and local discrete invariance, see Fig.~\ref{Fig1}: the global symmetry is exact in the absence of (relativistic) magnetocrystalline corrections, and the local invariance corresponds to the lattice spin Hamiltonian being invariant under the simultaneous reversal of a single spin and the signs of all exchange coupling constants defined on the bonds connecting this spin to all its neighbors.\cite{Mattis-PhysLettA1976,Villain-JPhysC1977,Toulouse-CommPhys1977} Note that these two symmetries are ultimately responsible for the strong degeneracy of the ground state, which presents no conventional long-range magnetic order (i.e, negligible macroscopic magnetization) and is rather described by a spin-spin correlator (namely, the Edwards-Anderson order parameter).

An effective (long-wavelength) description of frustrated magnets can be obtained by means of a renormalization procedure of the Kadanoff-Wilson type. In particular, volume average of the different physical quantities/operators is performed over mesoscopic length scales at which the magnetic system remains spatially homogeneous\cite{Volovik-ZhETF1978}. For instance, in the case of spin glasses, this length scale should be larger than the typical distance between minimal frustrated plaquettes but small compared to the size of magnetic disclinations \cite{Halperin-PRB1977,Andreev-ZhETF1978,Tserkovnyak-PRB2017}, whereas for the correlated spin glass phase of amorphous magnets, it should be larger than the magnetic correlation length of the system (which, in turn, is larger than the typical size of the crystal grains) \cite{Ochoa-PRB2018}. This coarse-graining promotes the local discrete invariance to a local exchange (gauge) invariance at macroscopic scales, namely, the localization of the (nonabelian) SO(3) group emerges as the order-parameter manifold of frustrated magnets \cite{Volovik-ZhETF1978}. A complete description of their magnetic sector also requires a secondary order parameter, namely the macroscopic spin density $\bm{m}(\vec{r},t)$.

Low-energy long-wavelength excitations around a local free-energy minimum are therefore described by a SO(3) field, $R(\vec{r},t)$, which parametrizes smooth and slowly varying rotations of the initial noncollinear ground state \cite{Halperin-PRB1977,Dombre-PRB1989,Ochoa-PRB2018}: given a reference (low-energy minimum) spin configuration $\{\bm{S}_{1i}\}_{i}$, we can define the local (averaged) spin-spin correlator \cite{Henley-AnnPhys1984}
\begin{equation}
\label{sscorr}
Q_{\alpha\beta}^{(2,1)}(\vec{r},t)=\sum_{i}w(\vec{r}_{i}-\vec{r})S_{2i}^{\alpha}(t)S_{1i}^{\beta}(t),
\end{equation} 
for an arbitrary spin configuration $\{\bm{S}_{2i}\}_{i}$ of the local minimum, where $w(\vec{x})$ is a nonnegative weight function (decreasing with $|\vec{x}|$) and $\vec{r}_{i}$ denotes the position of the $i$-th spin\cite{Footnote2}. Its singular value decomposition yields $Q^{(2,1)}=R_{L}DR_{R}^{\top}$, where $R_{L},\,R_{R}$ are orthogonal matrices and $D=\textrm{diag}(d_{1},d_{2},d_{3})$ is the matrix of singular values of the spin-spin correlator. The order parameter is then defined as $R=R_{L}R_{R}^{\top}$, which provides the optimal rotation matrix mapping the reference spin configuration onto $\{\bm{S}_{2i}\}_{i}$\cite{Footnote3}.

We note that magnetic disclinations appear as singularities in the order-parameter \cite{Footnote4}. However, rotations can always be locally defined as $\delta R_{\alpha\beta}=\epsilon_{\alpha\beta\gamma}\Omega^{\gamma}_{k}\delta x^{k}$,\cite{Dzyaloshinskii-AnnPhys1980} where $\epsilon_{\alpha\beta\gamma}$ is the Levi-Civita symbol and $\bm{\Omega}_{k}=\Omega^{\gamma}_{k}\bm{\hat{e}}_{\gamma}$, $k=x,y,z$, are Yang-Mills fields (in the spin space) describing the spatial variations of the collective spin rotation. The latter are defined as
\begin{equation}
\label{YM_fields}
\bm{\Omega}_{k}=\tfrac{i}{2}\text{Tr}\big[R|_{\textrm{loc}}^{\top}\bm{L}\,\partial_{k}R|_{\textrm{loc}}\big],
\end{equation}
where $[L^{\alpha}]_{\beta\gamma}=-i\epsilon_{\alpha\beta\gamma}$ are the generators of SO(3) and $R|_{\textrm{loc}}$ denotes any smooth and single-valued projection of the order parameter onto an open neighborhood of the point at which we evaluate the Yang-Mills fields. Similarly, the angular velocity of the order parameter reads  $\bm{\Omega}_{t}=i\,\text{Tr}\big[R|_{\textrm{loc}}^{\top}\bm{L}\,\partial_{t}R|_{\textrm{loc}}\big]/2$. 

The energetics of the aforementioned excitations are described by the O(4) nonlinear $\sigma$ model: in terms of the Yang-Mills fields and the secondary order parameter $\bm{m}$, the minimal energy model for frustrated magnets takes the form \cite{Volovik-ZhETF1978}
\begin{align}
\label{eq:energy_mag}
\mathcal{E}_{m}\big[R,\bm{m}\big]=\int d^{3}\vec{r}\left[\frac{A}{2}\vec{\bm{\Omega}}\bm{\odot}\vec{\bm{\Omega}}+\frac{\bm{m}^{2}}{2\chi}\right],
\end{align}
where $A$ and $\chi$ represent the order-parameter stiffness and the spin susceptibility of the system, respectively. Here, $\vec{\bm{\Omega}}=\bm{\Omega}_{k}\hat{e}_{k}$ and $\bm{\odot}$ denotes the scalar product in both real and spin space, i.e., $\vec{\bm{B}}\bm{\odot}\vec{\bm{C}}=B_{k}^{\alpha}C_{k}^{\alpha}$. Furthermore, we note that the (hydrodynamic) Poisson-bracket structure of frustrated magnets \cite{Dzyaloshinskii-AnnPhys1980} yields the constitutive relation $\bm{m}=\chi\bm{\Omega}_{t}$ in the absence of an external magnetic field, that is, a nonequilibrium spin density is dynamically generated in these magnetic platforms.

Unit-norm quaternions, $\mathbf{q}=(w,\bm{v})$, offer a convenient parametrization of SO(3) matrices \cite{Ochoa-PRB2018}. In general, the scalar component $w$ parametrizes the rotation angle and the vector component $\bm{v}$ lies along the rotation axis; for instance, by considering $w=\cos(\phi/2)$ and $\bm{v}=\sin(\phi/2)\,\bm{n}$, with $\phi$ and $\bm{n}$ being the local rotation angle and axis for spins, respectively, we retrieve the well-known Rodrigues' rotation formula, $R_{\alpha\beta}=\cos\phi\,\delta_{\alpha\beta}+(1-\cos\phi)n_{\alpha}n_{\beta}+\sin\phi\,\epsilon_{\alpha\gamma\beta} n_{\gamma}$. Unit-norm quaternions exhibit a group structure endowed by the Hamilton product $\mathbf{q}_{1}\wedge\mathbf{q}_{2}\equiv(w_{1}w_{2}-\bm{v}_{1}\bm{\cdot}\bm{v}_{2},w_{1}\bm{v}_{2}+w_{2}\bm{v}_{1}+\bm{v}_{1}\bm{\times}\bm{v}_{2})$ and the adjoint (inverse) operation $\mathbf{q}^{\star}\equiv(w,-\bm{v})$. It is worth remarking the usefulness of the Hamilton product in computing the product of rotation matrices, since $R_{1}\cdot R_{2}$ corresponds to $\mathbf{q}_{1}\wedge\mathbf{q}_{2}$. 

The identities $\bm{\Omega}_{t}=2\partial_{t}\mathbf{q}\wedge\mathbf{q}^{\star}$ and $\bm{\Omega}_{k}=2\partial_{k}\mathbf{q}\wedge\mathbf{q}^{\star}$ hold, so that the Yang-Mills fields play the role of the magnetization currents (in analogy with the case of conventional magnetism). Furthermore, we mention that, in the quaternion representation, the Lagrangian of the O(4) nonlinear $\sigma$ model describing the order-parameter dynamics of a frustrated magnet takes the following form
\begin{equation}
\label{eq:nlsigma}
\mathcal{L}=2\int d^{3}\vec{r}\,\Big[\chi\,\partial_{t}\mathbf{q}^{*}\wedge\partial_{t}\mathbf{q}-A\,\partial_{k}\mathbf{q}^{*}\wedge\partial_{k}\mathbf{q}\Big].
\end{equation}
A simple spin-wave analysis yields three independent Goldstone branches described by the sound velocity $c=\sqrt{A/\chi}$.\cite{Dasgupta-PRB2020} Since the unit-norm quaternion plays the role of the staggered order in a four-dimensional (order-parameter) space according to the above model, frustrated magnets can be conceived of as a \textit{generalized antiferromagnet}.\cite{Footnote5}

\section{Dynamics of itinerant carriers in a magnetically frustrated background}
\label{Dyn_itinerant_car}
The effective low-energy long-wavelength theory for itinerant carriers flowing within a frustrated magnet and being subjected to spin exchange with a textured magnetic background is embodied in the following Euclidean Lagrangian:
\begin{widetext}
\begin{equation}
\label{eq:Lag}
\mathcal{L}_{\textrm{E}}[x_{0},\vec{r};\Psi]=\hbar\Psi^{\dagger}\partial_{0}\Psi+\frac{\hbar^{2}}{2m_{\star}}\nabla\Psi^{\dagger}\cdot\nabla\Psi+g_{1}\bm{\Omega}_{\mu}\circ\big(i\Psi^{\dagger}\bm{\tau}\partial_{\mu}\Psi-i\partial_{\mu}\Psi^{\dagger}\bm{\tau}\Psi\big)-\Phi_{m}|\Psi|^{2}+\tfrac{\hbar}{2}\bm{\omega}_{m}\circ(\Psi^{\dagger}\bm{\tau}\Psi),
\end{equation}
\end{widetext}
where $\vec{r}$ denote dimensionless spatial coordinates (specific to the model considered, see Sec.~\ref{ref:Sec2DHA}), $x_{0}=it$ is the (Wick-rotated) imaginary time, $\{g_{k}\}_{k=0\ldots4}$ are coupling constants, and $\mu=x,y,z$, runs over spatial indices. $\Psi$ is the Fermi field describing the charge carrier and $\tau_{0}$, $\bm{\tau}$ denote the identity and the vector of Pauli matrices in the pseudospin space, respectively. This continuum theory can be built upon symmetry grounds and contains all possible terms up to second order in the spin density $\bm{m}$ and partial derivatives (of the Fermi field and the order parameter). There is a clear physical interpretation of the terms arising in the effective Hamiltonian: the first one (from left to right) is the usual kinetic term, parametrized in terms of an effective mass $m_{\star}$ for the conduction band of the carriers. The second term couples the spin current flowing within the frustrated magnet to the Yang-Mills fields $\bm{\Omega}_{x},\bm{\Omega}_{y}$ and $\bm{\Omega}_{z}$ describing the noncollinear spin background. As we will discuss in Sec.~\ref{STT}, this term yields some interesting spin-transfer physics because hopping of charge carriers favors a twist of the order parameter ($\vec{\bm{\Omega}}\neq\vec{\bm{0}}$) and viceversa.\cite{Shraiman-PRL1988} Furthermore, the third term stems from the coupling of the carrier density $\rho=\Psi^{\dagger}\Psi$ to an effective (spin texture-dependent) electric potential $\Phi_{m}=g_{2}\bm{m}^{2}+g_{3}\vec{\bm{\Omega}}\bm{\odot}\vec{\bm{\Omega}}$.\cite{Footnote6} The last term is responsible for the precessional spin dynamics of the charge carriers; the spin precession vector $\tfrac{\hbar}{2}\bm{\omega}_{m}=g_{0}\bm{\Omega}_{t}-g_{4}\bm{m}$ contains a (dynamical) Coriolis-type contribution, proportional to the angular velocity $\bm{\Omega}_{t}$ (note here that carriers flow in a rotating spin frame of reference adjusted to the spin background), and a spin-exchange contribution, resulting from the coupling of the background magnetization to the spin density $\bm{s}=\tfrac{\hbar}{2}\Psi^{\dagger}\bm{\tau}\Psi$ of itinerant carriers.

We devote Sec.~\ref{ref:Sec2DHA} to provide a formal microscopic derivation of this Lagrangian for one of the simplest models of frustrated magnetism, namely the two-dimensional Heisenberg antiferromagnet on a triangular lattice. Eq.~\eqref{eq:Lag} is built upon the kinetic term of the (single-band near half-filled) Hubbard Hamiltonian for conduction electrons, also defined on the triangular lattice, within the framework of the mean-field and slave-boson (SB) \cite{Kotliar-PRL1986,Li-PRB1989} approaches. It is worth remarking that, despite its simplicity, the doped Hubbard model captures the essential physics of high-temperature superconductivity and quantum magnetism. Furthermore, we also note that the SB formalism naturally accounts for the interplay between charge currents and spin textures, since it is founded on the idea that electron hopping in strongly correlated systems is accompanied by a backflow of spin excitations\cite{Kotliar-PRL1986}.

As a final remark, we mention that the above Lagrangian can be recast in the following Yang-Mills form for a SU(2) pseudospin \cite{Yang-PR1954}:
\begin{align}
\label{eq:YM_Lag}
\mathcal{L}_{\textrm{E}}[x_{0},\vec{r};\Psi]&=\hbar\Psi^{\dagger}\partial_{0}\Psi-\Phi_{m}|\Psi|^{2}+\tfrac{\hbar}{2}\bm{\omega}_{m}\circ(\Psi^{\dagger}\bm{\tau}\Psi)\\
&\hspace{0.2cm}-\frac{\hbar^{2}}{2m_{\star}}\Psi^{\dagger}\big(\partial_{\mu}-i g\,\bm{\Omega}_{\mu}\circ\bm{\tau}\big)^{2}\Psi,\nonumber
\end{align}
where we have neglected boundary terms, $g=2m_{\star}g_{1}/\hbar^{2}$ denotes a rescaled coupling constant, and $g_{3}$ is redefined in $\Phi_{m}$ as $g_{3}\equiv g_{3}+gg_{1}$. 

\section{Hydrodynamic theory for the itinerant charge and spin degrees of freedom}
\label{Hydro}
In this Section we construct hydrodynamic equations, based upon the Lagrangian~\eqref{eq:YM_Lag}, for the itinerant charge and spin densities of a frustrated magnet. We mention that an analogous derivation has been previously carried out in Ref.~\onlinecite{Jin-JPA2006} within the context of crystalline media subjected to weak spin-orbit interactions. To begin with, the saddle-point equation for the Fermi field, $\delta\mathcal{L}/\delta\Psi^{\dagger}=0$, reads
\begin{equation}
\label{eq:EoM_psi}
\hbar\partial_{0}\Psi-\Phi_{m}\Psi+(\tfrac{\hbar}{2}\bm{\omega}_{m}\circ\bm{\tau}\big)\Psi-\frac{\hbar^{2}}{2m_{\star}}\big(\partial_{\mu}-ig\,\bm{\Omega}_{\mu}\circ\bm{\tau}\big)^{2}\Psi=0,
\end{equation}
where, again, $g=2m_{\star}g_{1}/\hbar^{2}$. The equation of motion for $\Psi^{\dagger}$, $\delta\mathcal{L}/\delta\Psi=0$, becomes
\begin{align}
\label{eq:EoM_psi_exp2}
&\hbar\partial_{0}\Psi^{\dagger}=-\Phi_{m}\Psi^{\dagger}+\Psi^{\dagger}\big[\tfrac{\hbar}{2}\bm{\omega}_{m}-ig\tfrac{\hbar^{2}}{2m_{\star}}\partial_{\mu}\bm{\Omega}_{\mu}\big]\circ\bm{\tau}\\
&\hspace{0.3cm}-\frac{\hbar^{2}}{2m_{\star}}\partial_{\mu}^{2}\Psi^{\dagger}-ig\frac{\hbar^{2}}{m_{\star}}\partial_{\mu}\Psi^{\dagger}\big(\bm{\Omega}_{\mu}\circ\bm{\tau}\big)+\frac{\hbar^{2}g^{2}}{2m_{\star}}\vec{\bm{\Omega}}\bm{\odot}\vec{\bm{\Omega}}\,\Psi^{\dagger}.\nonumber
\end{align}
Second, the hydrodynamic equation for the itinerant spin density can be obtained via the linear combination $\big(\Psi^{\dagger}\frac{\hbar}{2}\bm{\tau}\big)\textrm{Eq.}~\eqref{eq:EoM_psi}-\textrm{Eq.}~\eqref{eq:EoM_psi_exp2}\big(\frac{\hbar}{2}\bm{\tau}\hspace{0.05cm}\Psi\big)$. It reads
\begin{align}
\label{eq:hydro_eq_spin}
\partial_{t}\bm{s}+\partial_{\mu}\bm{J}_{\mu}=\bm{\omega}_{m}\bm{\times}\bm{s}+2g\,\bm{J}_{\mu}\bm{\times}\bm{\Omega}_{\mu}.
\end{align}
Similarly, by combining the above saddle-point equations in the form $\Psi^{\dagger}\textrm{Eq.}~\eqref{eq:EoM_psi}-\textrm{Eq.}~\eqref{eq:EoM_psi_exp2}\Psi$, we also derive the following continuity equation for the probability density:
\begin{equation}
\label{eq:hydro_eq_prob}
\partial_{t}\rho+\partial_{\mu}j_{\mu}=0.
\end{equation}
These hydrodynamic equations have been obtained with account of the identities $\big(\bm{A}\circ\bm{\tau}\big)\bm{\tau}=\bm{A}\tau_{0}-i\bm{A}\bm{\times}\bm{\tau}$ and $\bm{\tau}\big(\bm{A}\circ\bm{\tau}\big)=\bm{A}\tau_{0}+i\bm{A}\bm{\times}\bm{\tau}$ in the spin space, which can be inferred from the algebra of Pauli matrices. A complete hydrodynamic description of the itinerant degrees of freedom requires the following constitutive relations for the probability and spin currents:
\begin{align}
\label{prob_curr}
j_{\mu}&=\frac{\hbar}{2m_{\star}}\big(i\partial_{\mu}\Psi^{\dagger}\Psi-i\Psi^{\dagger}\partial_{\mu}\Psi\big)-\frac{4g_{1}}{\hbar^{2}}\bm{\Omega}_{\mu}\circ\bm{s},\\
\label{spin_curr}
\bm{J}_{\mu}&=\frac{\hbar^{2}}{4m_{\star}}\big(i\partial_{\mu}\Psi^{\dagger}\bm{\tau}\Psi-i\Psi^{\dagger}\bm{\tau}\partial_{\mu}\Psi\big)-g_{1}\rho\,\bm{\Omega}_{\mu}.
\end{align}
New contributions to the thermodynamic fluxes, ascribed to the emergent coupling between the itinerant spin and magnetization currents, therefore appear. These unconventional terms, embodying the spin exchange of itinerant carriers with the noncollinear magnetic background, are linear with the Yang-Mills fields: the probability current depends on the projection of $\vec{\bm{\Omega}}$ onto the itinerant spin polarization, whereas the spin current depends on the product of the Yang-Mills fields with the density of carriers. As we will discuss in the next Section, these dependences underlie the spin-transfer response of the electron fluid. Furthermore, we note the presence of source terms in the right hand side of Eq.~\eqref{eq:hydro_eq_spin}, acting as magnetic torques on the itinerant spins: the first term describes the usual spin precession of the itinerant carriers under the action of the 'external field' $\bm{\omega}_{m}$; the second one, with no analogy in conventional magnetism, also emerges from the coupling between the magnetization and itinerant spin currents, and is nonzero when the latter are noncollinear. We also mention that the projection of the spin current $\vec{\bm{J}}$ onto the direction of the spin density $\bm{s}$ reads
\begin{align}
\label{spin_curr_proj}
\left(\Psi^{\dagger}\bm{\tau}\Psi\right)\circ\bm{J}_{\mu}&=\frac{\hbar}{2e}\rho j_{\mu}^{e},\nonumber
\end{align}
namely, the spin current projected along the direction of the spin density equals the charge current times the density of carriers (up to the charge-to-spin conversion factor).

\section{Spin-transfer torques}
\label{STT}

We start this Section by mentioning that $\{\rho,\bm{s}\}$ are the thermodynamic variables describing the fermion fluid, whereas $\big\{\vec{j},\vec{\bm{J}}\hspace{0.03cm}\big\}$ represent the conjugated thermodynamic fluxes. In terms of these quantities, we can recast the Hamiltonian part of the Euclidean Lagrangian~\eqref{eq:Lag} as
\begin{widetext}
\begin{equation}
\label{eq:energy_carriers}
\mathcal{E}_{c}[\rho,\bm{s}]=\int d\vec{r}\,\bigg[\tfrac{1}{2m_{\star}\rho}\big(\nabla\bm{s}\big)^{2}+\tfrac{m_{\star}}{2\rho}\left(j_{\mu}+\tfrac{2g}{m_{\star}}\bm{\Omega}_{\mu}\circ\bm{s}\right)^{2}-2g\,\vec{\bm{\Omega}}\bm{\odot}\vec{\bm{J}}-\big(\Phi_{m}+2gg_{1}\vec{\bm{\Omega}}\bm{\odot}\vec{\bm{\Omega}}\big)\rho+\bm{\omega}_{m}\circ\bm{s}\bigg].
\end{equation}
\end{widetext}
Here, we have considered the identity
\begin{equation}
\label{eq:grad_psi}
\big(\partial_{\mu}\bm{s}\big)^{2}=-\big[m_{\star}j_{\mu}+2g\,\bm{\Omega}_{\mu}\circ\bm{s}\big]^{2}+\hbar^{2}\rho\,\partial_{\mu}\Psi^{\dagger}\partial_{\mu}\Psi,
\end{equation}
for the derivatives of the itinerant spin density. This expression can be derived directly with account of the relation $\bm{\tau}_{\alpha\beta}\circ\bm{\tau}_{\gamma\delta}=2\delta_{\alpha\delta}\delta_{\beta\gamma}-\delta_{\alpha\beta}\delta_{\gamma\delta}$ for the product of Pauli matrices and the constituent relation~\eqref{prob_curr} for the probability current. The total energy $\mathcal{E}[\rho,\bm{s},R,\bm{m}]$ of the frustrated magnet consists of the above energy as well as the energy of the magnetic sector, see Eq.~\eqref{eq:energy_mag}. The thermodynamic forces conjugated to the spin degrees of freedom are thus
\begin{align}
\label{eq:thermo_forces}
&\bm{f}_{\bm{m}}=-\tfrac{\delta\mathcal{E}}{\delta\bm{m}}=(2g_{2}\rho-\chi^{-1})\bm{m}+\tfrac{2(g_{4}-g_{0}/\chi)}{\hbar}\bm{s},\\
&\bm{f}_{\bm{s}}=-\tfrac{\delta\mathcal{E}}{\delta\bm{s}}=\tfrac{\nabla^{2}\bm{s}}{m_{\star}\rho}-\tfrac{2g}{\rho}\left[\vec{j}+\tfrac{2g}{m_{\star}}\vec{\bm{\Omega}}\circ\bm{s}\right]\cdot\vec{\bm{\Omega}}-\bm{\omega}_{m},
\end{align}
from which we can recast the itinerant spin density and the magnetization of the system as $\bm{s}=\frac{\hbar}{2g_{4}}\bm{f}_{\bm{m}}+\ldots$ and $\bm{m}=\frac{\hbar}{2g_{4}}\bm{f}_{\bm{s}}+\ldots$ in terms of the thermodynamic currents. Here, $'\ldots'$ denotes other terms irrelevant to our discussion and we have taken into account the constitutive relation $\bm{m}=\chi\bm{\Omega}_{t}$ for the localized macroscopic spin density as well as the fact that $g_{0}/g_{4}\chi\ll1$. Furthermore, in the absence of a magnetic texture $(\vec{\bm{\Omega}}\equiv\vec{\bm{0}})$, Eq.~\eqref{eq:hydro_eq_prob} yields a conservation law for the electric charge, $\partial_{t}(e\rho)+\partial_{\mu}j^{e}_{\mu}=0$, with
\begin{equation}
\label{eq:charge_curr}
j^{e}_{\mu}\equiv ej_{\mu}=\frac{\hbar e}{2m_{\star}}\big(i\partial_{\mu}\Psi^{\dagger}\Psi-i\Psi^{\dagger}\partial_{\mu}\Psi\big)=\vartheta_{\mu\mu_{1}}E_{\mu_{1}}.
\end{equation}
The last identity arises from linear response theory, where $\hat{\vartheta}$ and $\vec{E}$ denote the conductivity tensor (assumed to be symmetric, namely, purely dissipative) and the driving electric field, respectively. In the general case, the charge current reads 
\begin{equation}
\label{eq:charge_curr2}
\vec{j}^{e}=\hat{\vartheta}\vec{E}-\frac{4eg_{1}}{\hbar^{2}}\vec{\bm{\Omega}}\circ\bm{s}=\hat{\vartheta}\vec{E}-\frac{2eg_{1}}{\hbar g_{4}}\vec{\bm{\Omega}}\circ\bm{f}_{\bm{m}}+\ldots
\end{equation}
Thermodynamic fluxes and forces in the spin and charge sectors are related by the following Onsager matrix \cite{Zarzuela-PRB2019}:
\begin{align}
\label{eq:Onsager_matrix}
\left(\begin{array}{c}
\partial_t\bm{m}\\
\vec{j}^{e}
\end{array}\right)=\left(\begin{array}{ccc}
\cdot\star\cdot & & \hat{L}_{\textrm{sq}} \\
 \hat{L}_{\textrm{qs}} & & \hat{\vartheta}
\end{array}\right)
\left(\begin{array}{c}
\bm{f}_{\bm{m}}\\
\vec{E}
\end{array}\right),
\end{align}
where $\cdot\star\cdot$ denotes a linear-response coefficient inconsequential for our discussion. From Eq.~\eqref{eq:charge_curr2} we identify $\hat{L}_{\textrm{qs}}\big|_{\mu\mu_{1}}=-\frac{2eg_{1}}{\hbar g_{4}}\Omega_{\mu}^{\mu_{1}}$ and, since off-diagonal blocks in the above Onsager matrix are related by the reciprocal relation $\hat{L}_{\textrm{sq}}=-\hat{L}_{\textrm{qs}}^{\top}$, we obtain the following contribution (to the lowest order in the Yang-Mills fields) to the magnetic torque acting on the frustrated magnet:
\begin{align}
\label{eq:mag_torque1}
\partial_{t}\bm{m}&=\frac{2eg_{1}}{\hbar\vartheta g_{4}}\vec{j}^{e}\cdot\vec{\bm{\Omega}},
\end{align}
where, for the sake of simplicity, we have assumed a diagonal structure for the conductivity tensor. We note in passing that this expression for the dissipative component of the magnetic torque has been suggested based on symmetry grounds in Ref.~\onlinecite{Zarzuela-PRB2019}. Similarly, we have also found a second contribution to the spin-transfer torque of the form
\begin{equation}
\label{eq:mag_torque2}
\partial_{t}\bm{m}=\frac{8e^{2}g_{1}^{2}}{\hbar^{3}\vartheta g_{4}}(\vec{\bm{\Omega}}\circ\bm{s})\cdot\vec{\bm{\Omega}},
\end{equation} 
which is second order in the Yang-Mills fields and depends on the spin polarization of the itinerant carriers; in particular, this magnetic torque vanishes when the itinerant spin density is orthogonal to the Yang-Mills fields. Furthermore, the corresponding prefactors depend on the spin-to-charge conversion factor and the ratio of the coupling constant $g_{1}$ to $g_{4}$. Therefore, we conclude that the spin-transfer physics of a frustrated magnet is mediated by the interplay between two interactions, namely the effective exchange between the itinerant and localized spin densities, described by the term $\bm{\omega}_{\bm{m}}\circ\bm{s}$ in Eq.~\eqref{eq:energy_carriers}, and the emergent coupling between the itinerant spin and magnetization currents, described by the term $\propto\vec{\bm{\Omega}}\bm{\odot}\vec{\bm{J}}$ in the same equation.

\section{Microscopic derivation of the effective Lagrangian for itinerant carriers}
\label{ref:Sec2DHA}

We present in this Section a microscopic derivation of the Lagrangian \eqref{eq:Lag} for the two-dimensional Heisenberg antiferromagnet on a triangular lattice. Since this simple model belongs to the universality class of frustrated magnets dominated by isotropic exchange \cite{Dombre-PRB1989,Azaria-PRL1992,Chubukov-PRL1994}, we expect the resulting low-energy long-wavelength Lagrangian to also describe the dynamics of itinerant carriers in the general case. Our starting point is the single-band Hubbard model for electrons at near half filling
\begin{equation}
\label{eq:Hubbard}
H=\sum_{\langle i,j\rangle,\sigma}t_{ij}c_{i,\sigma}^{\dagger}c_{j,\sigma}+U\sum_{i}n_{i,\uparrow}n_{i,\downarrow},
\end{equation}
where $\{i,j\}$ run over nearest neighbors (NN), $t_{ij}$ are NN hopping matrix elements, $U$ represents the strength of the on-site Coulomb repulsion and $\sigma=\uparrow,\downarrow$ denotes both projections of the electron spin onto the quantization axis. $c_{i,\sigma}$ and $n_{i,\sigma}$ are the electron annihilation operator and the occupation number for the spin projection $\sigma$, respectively, at the $i$-th site. We will focus primarily on the tight-binding term of the Hubbard Hamiltonian and use the spin-rotation invariant (SRI) form of the slave boson (SB) representation for electron operators \cite{Li-PRB1989}. The SB formulation of strongly correlated systems, which has been widely applied to describe the physics of high-$T_{\textrm{c}}$ superconductors and heavy fermion compounds, is rooted in the idea that electron hopping is accompanied by a \emph{backflow} of spin excitations \cite{Kotliar-PRL1986}. The latter makes it apposite to incorporate naturally the exchange between localized and itinerant spin degrees of freedom into a low-energy transport theory for frustrated magnets. Within the SRI SB representation, the electron operators are cast as (see Appendix A for further details)
\begin{equation}
\label{eq:SRI-SB}
c_{i,\sigma}=\sum_{\sigma'}z_{i,\sigma\sigma'}f_{i,\sigma'},
\end{equation}
in terms of (spinless) pseudofermion operators $f_{i,\sigma}$. The operator matrix $\underline{z_{i}}$ is defined as
 \begin{align}
 \label{eq:matrix_z}
 \underline{z_{i}}&=[(1-d_{i}^{\dagger}d_{i})\tau_{0}-\underline{S_{i}}^{\dagger}\underline{S_{i}}]^{-1/2}(e_{i}^{\dagger}\underline{S_{i}}+\underline{S_{T,i}}^{\dagger}d_{i})\\
 &\hspace{0.8cm}\times[(1-e_{i}^{\dagger}e_{i})\tau_{0}-\underline{S_{T,i}}^{\dagger}\underline{S_{T,i}}]^{-1/2},\nonumber
 \end{align}
 with the operators $\underline{S_{i}}$ and $\underline{S_{T,i}}=\hat{T}\underline{S_{i}}\hat{T}^{-1}$ (time-reversed spin operator) being given by
 \begin{align}
 \underline{S_{i}}&=\tfrac{1}{\sqrt{2}}\left[s_{0,i}\tau_{0}+\bm{s}_{i}\circ\bm{\tau}\right],\\
 \underline{S_{T,i}}&=\tfrac{1}{\sqrt{2}}\left[s_{0,i}\tau_{0}-\bm{s}_{i}\circ\bm{\tau}\right].
 \end{align}
 Here, $\{e_{i},d_{i},s_{0,i},\bm{s}_{i}\}$ are SB operators accounting for empty, double and single occupied sites, respectively. With account of the constraint $e_{i}^{\dagger}e_{i}+s_{0,i}^{\dagger}s_{0,i}+\bm{s}_{i}\hspace{0.01cm}^{\dagger}\circ\bm{s}_{i}+d_{i}^{\dagger}d_{i}=1$ on the occupancy of the sites, we obtain the identity
 \begin{align}
 \label{eq:int_step1}
 \underline{S_{i}}^{\dagger}\underline{S_{i}}&=\tfrac{1}{2}\big[
 (s_{0,i}^{\dagger}\bm{s}_{i}+\bm{s}_{i}\hspace{0.01cm}^{\dagger}s_{0,i}+i\bm{s}_{i}\hspace{0.01cm}^{\dagger}\hspace{-0.05cm}\bm{\times}\bm{s}_{i})\circ\bm{\tau}\\
 &\hspace{1.0cm}+(1-d_{i}^{\dagger}d_{i}-e_{i}^{\dagger}e_{i})\tau_{0}\big].\nonumber
 \end{align}
In the spirit of Refs.~\onlinecite{Li-PRB1989,Fresard-EPL1991} and~\onlinecite{Fresard-JCM1992}, we assume that the spatial dependence of the mean-field (slave) Bose fields is dictated by that of the background spin field: quantities transforming as scalars in the spin space (namely, $e_{i},d_{i}$ and $s_{0,i}$) will be considered as spatially homogeneous hereafter, and those transforming as vectors (namely, $\bm{s}_{i}$) will adjust to the magnetic texture adiabatically. This leads to $\langle e_{i}^{\dagger}\rangle_{\textrm{MF}}=\langle e_{i}\rangle_{\textrm{MF}}\equiv e$, $\langle d_{i}^{\dagger}\rangle_{\textrm{MF}}=\langle d_{i}\rangle_{\textrm{MF}}\equiv d$, $\langle s_{0,i}^{\dagger}\rangle_{\textrm{MF}}=\langle s_{0,i}\rangle_{\textrm{MF}}\equiv s_{0}$ and $\langle \bm{s}_{i}\hspace{0.02cm}^{\dagger}\rangle_{\textrm{MF}}=\langle \bm{s}_{i}\rangle_{\textrm{MF}}\equiv S\bm{n}(\vec{r}_{i})$, where $\bm{n}$ denotes the unit-norm spin field describing the background magnetic texture and $S$ denotes an effective value of the spins. Therefore, with account of Eq.~\eqref{eq:int_step1} the operator matrix \eqref{eq:matrix_z} reads
\begin{widetext}
 \begin{align}
 \label{eq:matrix_z_MF}
 \langle\underline{z_{i}}\rangle_{\textrm{MF}}&=\sqrt{2}\Big[\tau_{0}-\frac{2s_{0}S}{1+e^{2}-d^{2}}\bm{n}(\vec{r}_{i})\circ\bm{\tau}\Big]^{-1/2}\frac{(e+d)s_{0}\tau_{0}+(e-d)S\bm{n}(\vec{r}_{i})\circ\bm{\tau}}{\sqrt{1+e^{2}-d^{2}}\sqrt{1+d^{2}-e^{2}}}\Big[\tau_{0}+\frac{2s_{0}S}{1+d^{2}-e^{2}}\bm{n}(\vec{r}_{i})\circ\bm{\tau}\Big]^{-1/2}.
 \end{align}
 \end{widetext}
 By invoking Sylvester's formula in the form
 \begin{equation}
 \label{eq:Sylvester}
 f\big(a\,\bm{\xi}\circ\bm{\tau}\big)=\frac{f(a)+f(-a)}{2}\tau_{0}+\frac{f(a)-f(-a)}{2}\bm{\xi}\circ\bm{\tau},
\end{equation}
where $\bm{\xi}$ is a unit-norm spin vector and $f$ is an analytical function at $a$ and $-a$, we can recast the above equation as $\big[\textrm{we use }f_{1}(x)=(1-x)^{-1/2}$, $a_{1}=\frac{2s_{0}S}{1+e^{2}-d^{2}}$ and $f_{2}(x)=(1+x)^{-1/2}$, $a_{2}=\frac{2s_{0}S}{1+d^{2}-e^{2}}\big]$:
\begin{widetext}
\begin{align}
\label{eq:matrix_z_MF2}
 \langle\underline{z_{i}}\rangle_{\textrm{MF}}&=\frac{1}{2\sqrt{2}}\Big(\big[a(+,-)+a(+,+)\big]\tau_{0}+\big[a(+,-)-a(+,+)\big]\bm{n}(\vec{r}_{i})\circ\bm{\tau}\Big)\Big([e+d]s_{0}\tau_{0}+[e-d]S\bm{n}(\vec{r}_{i})\circ\bm{\tau}\Big)\\
 &\hspace{2cm}\times\Big(\big[a(-,+)+a(-,-)\big]\tau_{0}+\big[a(-,+)-a(-,-)\big]\bm{n}(\vec{r}_{i})\circ\bm{\tau}\Big)\nonumber,
\end{align}
\end{widetext}
where we have introduced the auxiliary function $a(\sigma_{1},\sigma_{2})\equiv\big[1+\sigma_{1}(e^{2}-d^{2})+2\sigma_{2}s_{0}S\big]^{-1/2}$. By expanding the above product and casting all terms on the basis $\{\tau_{0},\bm{\tau}\}$, we obtain the expression
\begin{align}
\label{eq:matrix_z_MF3}
 \langle\underline{z_{i}}\rangle_{\textrm{MF}}&=A_{1}(s_{0},e,d)\tau_{0}+A_{2}(s_{0},e,d)\,\bm{n}(\vec{r}_{i})\circ\bm{\tau},
\end{align}
for the mean-field value of the matrix $\underline{z_{i}}$, where the coefficients read\cite{Fresard-EPL1991,Fresard-JCM1992}
\begin{widetext}
\begin{align}
\label{eq:A1}
A_{1}(s_{0},e,d)&\equiv\frac{1}{\sqrt{2}}\Big(s_{0}(e+d)\big[a(+,-)a(-,+)+a(+,+)a(-,-)\big]+S(e-d)\big[a(+,-)a(-,+)-a(+,+)a(-,-)\big]\Big),\\
\label{eq:A2}
 A_{2}(s_{0},e,d)&\equiv\frac{1}{\sqrt{2}}\Big(s_{0}(e+d)\big[a(+,-)a(-,+)-a(+,+)a(-,-)\big]+S(e-d)\big[a(+,-)a(-,+)+a(+,+)a(-,-)\big]\Big).
\end{align}
\end{widetext}
The tight-binding term of the Hubbard Hamiltonian in the SRI SB representation becomes
\begin{align}
\label{eq:tb_hamiltonian}
H_{\textrm{tb}}^{\textrm{SRI SB}}&=\sum_{\langle i,j\rangle,\sigma_{1},\sigma_{2}}t_{ij}(f_{i,\sigma_{1}}^{\dagger}z_{i,\sigma_{1}\sigma}^{\dagger})(z_{j,\sigma\sigma_{2}}f_{j,\sigma_{2}})\\
&=\sum_{\langle i,j\rangle}t_{ij}\Psi_{i}^{\dagger}\big(\underline{z_{i}}^{\dagger}\underline{z_{j}}\big)\Psi_{j},\nonumber
\end{align}
 where $\Psi_{i}$ denotes the (pseudo)spinor describing the itinerant carrier at the $i$-th site $(\Psi_{i,\sigma}\equiv f_{i,\sigma})$. Furthermore, within the mean-field approach discussed above, we obtain the following expression for the matrix product: 
 \begin{align}
 \label{eq:matrix_z_prod}
 \langle\underline{z_{i}}^{\dagger}\underline{z_{j}}\rangle_{\textrm{MF}}&=\langle\underline{z_{i}}^{\dagger}\rangle_{\textrm{MF}}\langle\underline{z_{j}}\rangle_{\textrm{MF}}=\Big[A_{1}^{2}+A_{2}^{2}\,\bm{n}(\vec{r}_{i})\circ\bm{n}(\vec{r}_{j})\Big]\tau_{0}\nonumber\\
 &\hspace{-1.6cm}+\Big[A_{1}A_{2}\big(\bm{n}\big(\vec{r}_{i})+\bm{n}(\vec{r}_{j})\big)+iA_{2}^{2}\,\bm{n}(\vec{r}_{i})\bm{\times}\bm{n}(\vec{r}_{j})\Big]\circ\bm{\tau}.
 \end{align}
 
 \begin{figure}
\begin{center}
\includegraphics[width=1.0\columnwidth]{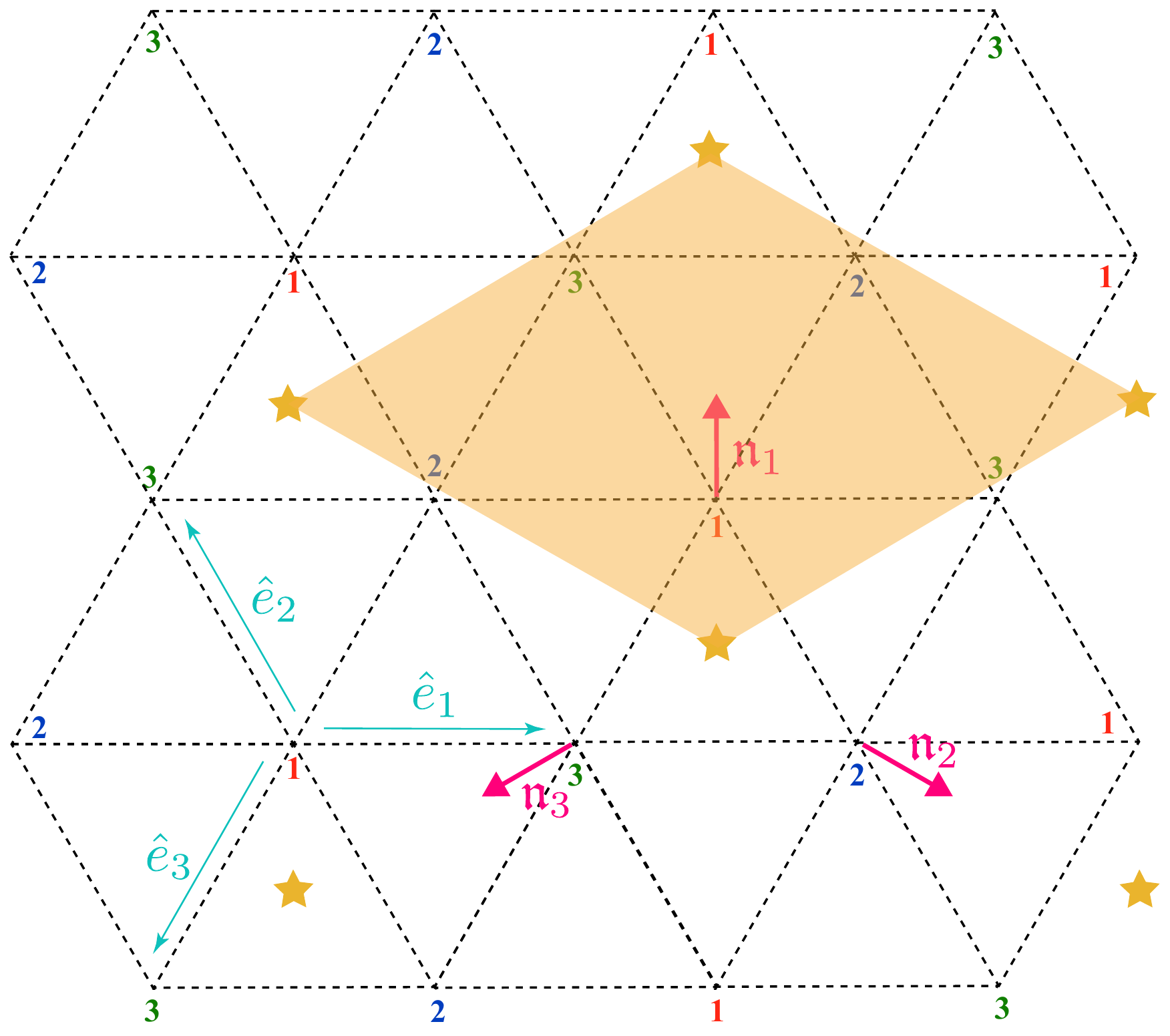}
\caption{Triangular lattice for the Heisenberg antiferromagnet. The three sublattices are labelled by the numbers 1,2 and 3. The reference ordered state for spins is represented by magenta arrows. Director vectors along the bonds are depicted by turquoise arrows. Golden stars denote the magnetic lattice of the system and the peach rhombus depicts the magnetic unit cell.}
\vspace{-0.5cm} 
\label{Fig2}
\end{center}
\end{figure}
 
The classical ground state of the triangular antiferromagnet can be described by means of a single upright triangular plaquette, which repeats itself along the lattice \cite{Dombre-PRB1989}: in each plaquette, whose vertices belong to three different sublattices, each spin is oriented at an angle $2\pi/3$ with respect to the other two, rotating counterclockwise from one to the next, see Fig.~\ref{Fig2}. Within a given sublattice, the orientation of the spins may change between neighboring plaquettes. A reference ordered state for spins in the plaquettes is chosen, see Fig.~\ref{Fig2}, which is described by the director vectors $\frak{n}_{1}=(0,1,0)^{\top}$, $\frak{n}_{2}=(\sqrt{3}/2,-1/2,0)^{\top}$ and $\frak{n}_{3}=(-\sqrt{3}/2,-1/2,0)^{\top}$ on the three sublattices. Following Ref.~\onlinecite{Dombre-PRB1989}, the lattice sites will be indexed by means of two subindices $(k,\lambda)$, where $k$ and $\lambda$ denote the plaquette number and sublattice number (within the plaquette), respectively. According to this indexation, the unit-norm spins can be casted as \cite{Dombre-PRB1989}
\begin{equation}
\label{eq:spins}
\bm{n}^{(\lambda)}(\vec{r}_{k})=\frac{R(\vec{r}_{k})[\frak{n}_{\lambda}+a\bm{N}(\vec{r}_{k})]}{\sqrt{1+2a\,\frak{n}_{\lambda}\circ\bm{N}(\vec{r}_{k})+a^{2}\bm{N}^{2}(\vec{r}_{k})}},
\end{equation}
where $a$ denotes the nearest-neighbor distance and $\bm{N}$ is a vector related to the net magnetization of the system. We note that we have disregarded the time dependence of the rotation matrix $R$ in Eq.~\ref{eq:spins} for the sake of simplicity. The expansion of the above equation to the lowest order in $a|\bm{N}|\ll1$ yields the relations
\begin{align}
\label{eq:spin_expansion1}
\hspace{-0.5cm}\bm{n}^{(\lambda)}(\vec{r}_{k})&\simeq R(\vec{r}_{k})\big[\frak{n}_{\lambda}+a\{\bm{N}(\vec{r}_{k})-(\bm{N}(\vec{r}_{k})\hspace{-0.05cm}\circ\hspace{-0.05cm}\frak{n}_{\lambda})\frak{n}_{\lambda}\}\big],\\
\label{eq:spin_expansion2}
\bm{m}(\vec{r}_{k})&=S\sum_{\lambda=1}^{3}\bm{n}^{(\lambda)}(\vec{r}_{k})\simeq3aSR(\vec{r}_{k})[\hat{T}\bm{N}(\vec{r}_{k})],
\end{align}
where $\hat{T}$ is a tensor whose components are defined per $T_{\alpha\beta}=\delta_{\alpha\beta}-\frac{1}{3}\sum_{\lambda=1}^{3}\frak{n}_{\lambda}^{\alpha}\frak{n}_{\lambda}^{\beta}$ ($T_{xx}=T_{yy}=1/2; T_{zz}=1; T_{\alpha\beta}=0$ otherwise). Therefore, $\bm{N}(\vec{r}_{k})$ is a spin vector parametrizing the total magnetization of the $k$-th plaquette.

In what follows, we will expand Eqs.~\eqref{eq:tb_hamiltonian} and~\eqref{eq:matrix_z_prod} by incorporating the above parametrization for the background spin field. However, we will redefine $R$ and $\bm{N}$ on each site rather than on each plaquette; the underlying reason is that, in doing so, the physical meaning of the fields becomes less obvious but the results derived are equivalent and the procedure is simpler \cite{Dombre-PRB1989}. Therefore, Eq.~\eqref{eq:tb_hamiltonian} can be recast as
\begin{widetext}
\begin{align}
\label{eq:tb_hamiltonian2}
H_{\textrm{tb}}^{\textrm{SRI SB}}=\frac{1}{2}\sum_{\lambda=1}^{3}\sum_{i\in\Delta_{\lambda}}\sum_{\eta\neq\lambda}\sum_{j\in\Delta_\eta\cap\textrm{NN}(i)} t_{ij}\Psi_{i}^{(\lambda)}\hspace{0.01cm}^{\dagger}\big(\langle\underline{z_{i}}^{(\lambda)}\rangle_{\textrm{MF}}^{\dagger}\langle\underline{z_{j}}^{(\eta)}\rangle_{\textrm{MF}}\big)\Psi_{j}^{(\eta)} +\textrm{h.c.},
\end{align}
\end{widetext}
where $\lambda,\eta$ are sublattice indices and $\Delta_{\rho}$ denotes the $\rho$-th sublattice. The meaning of this expression is the following: first, a spin of a given sublattice (index $\lambda$) is considered, which interacts with its six nearest neighbors belonging to the two other sublattices (index $\eta$). Second, the sum is carried over the three sublattices and then divided by $2$ to avoid the double counting of the terms \cite{Footnote7}.

Since our interest lies in the continuum limit of the tight-binding Hubbard Hamiltonian, we expand next the spin field $\bm{n}^{(\eta)}(\vec{r}_{j})$ and the Fermi field $\Psi^{(\eta)}(\vec{r}_{j})$ around the lattice position $\vec{r}_{i}$:
\begin{widetext}
\begin{align}
\label{eq:expansions}
\bm{n}^{(\eta)}(\vec{r}_{j})&\simeq\bm{n}^{(\eta)}(\vec{r}_{i})+a(\hat{e}_{ij}^{(\lambda,\eta)}\cdot\nabla)\bm{n}^{(\eta)}(\vec{r}_{i})+\frac{1}{2}a^{2}(\hat{e}_{ij}^{(\lambda,\eta)}\cdot\nabla)^{2}\bm{n}^{(\eta)}(\vec{r}_{i})+\ldots,\\
\Psi^{(\eta)}(\vec{r}_{j})&\simeq\Psi(\vec{r}_{i})+a(\hat{e}_{ij}^{(\lambda,\eta)}\cdot\nabla)\Psi(\vec{r}_{i})+\frac{1}{2}a^{2}(\hat{e}_{ij}^{(\lambda,\eta)}\cdot\nabla)^{2}\Psi(\vec{r}_{i})+\ldots,\nonumber
\end{align}
\end{widetext}
where $\hat{e}_{ij}^{(\lambda,\eta)}=\epsilon_{\eta\lambda\rho}\hat{e}_{k(i,j)}$ are director vectors defined along the bonds between NN, see Fig.~\ref{Fig2}. Here, $\{\lambda,\eta,\rho\}=\{1,2,3\}$, $\hat{e}_{1}=(1,0)$, $\hat{e}_{2}=(-1,\sqrt{3})/2$, $\hat{e}_{3}=-(1,\sqrt{3})/2$ and $k:\Delta_{\lambda}\times\Delta_{\eta}\rightarrow\{1,2,3\}$ is an index function depending solely on the relative distance $\vec{r}_{j}-\vec{r}_{i}$. Consequently, the following expansions hold up to second order in the nearest-neighbor distance:
\begin{widetext}
\begin{align}
\label{eq:expansions2}
\bm{n}^{(\lambda)}(\vec{r}_{i})+\bm{n}^{(\eta)}(\vec{r}_{j})&\simeq \big[\bm{n}^{(\lambda)}+\bm{n}^{(\eta)}\big](\vec{r}_{i})+a\big(\hat{e}_{ij}^{(\lambda,\eta)}\cdot\nabla\big)\bm{n}^{(\eta)}(\vec{r}_{i})+\frac{1}{2}a^{2}(\hat{e}_{ij}^{(\lambda,\eta)}\cdot\nabla)^{2}\bm{n}^{(\eta)}(\vec{r}_{i}),\\
\bm{n}^{(\lambda)}(\vec{r}_{i})\circ\bm{n}^{(\eta)}(\vec{r}_{j})&\simeq\big[\bm{n}^{(\lambda)}\circ\bm{n}^{(\eta)}\big](\vec{r}_{i})+a\,\bm{n}^{(\lambda)}(\vec{r}_{i})\circ\big(\hat{e}_{ij}^{(\lambda,\eta)}\cdot\nabla\big)\bm{n}^{(\eta)}(\vec{r}_{i})+\frac{1}{2}a^{2}\bm{n}^{(\lambda)}(\vec{r}_{i})\circ(\hat{e}_{ij}^{(\lambda,\eta)}\cdot\nabla)^{2}\bm{n}^{(\eta)}(\vec{r}_{i}),\nonumber\\
\bm{n}^{(\lambda)}(\vec{r}_{i})\bm{\times}\bm{n}^{(\eta)}(\vec{r}_{j})&\simeq\big[\bm{n}^{(\lambda)}\bm{\times}\bm{n}^{(\eta)}\big](\vec{r}_{i})+a\,\bm{n}^{(\lambda)}(\vec{r}_{i})\bm{\times}\big(\hat{e}_{ij}^{(\lambda,\eta)}\cdot\nabla\big)\bm{n}^{(\eta)}(\vec{r}_{i})+\frac{1}{2}a^{2}\bm{n}^{(\lambda)}(\vec{r}_{i})\bm{\times}(\hat{e}_{ij}^{(\lambda,\eta)}\cdot\nabla)^{2}\bm{n}^{(\eta)}(\vec{r}_{i}).\nonumber
\end{align}
\end{widetext}
By incorporating Eqs.~\eqref{eq:expansions} and~\eqref{eq:expansions2} into Eqs.~\eqref{eq:matrix_z_prod} and~\eqref{eq:tb_hamiltonian2} we obtain, after some algebra, the following expression for the tight-binding Hamiltonian in the SRI SB representation
\begin{widetext}
\begin{align}
\label{eq:tb_hamiltonian_LW}
\langle H_{\textrm{tb}}^{\textrm{SRI SB}}\rangle_{\textrm{MF}}\simeq-t\sum_{k\in\Delta_{c}}&\Psi^{\dagger}\Bigg\{\frac{3}{2}\big[3(2A_{1}^{2}-A_{2}^{2})+(A_{2}/S)^{2}\bm{m}^{2}\big]\tau_{0}\Psi+6(A_{1}A_{2}/S)\bm{m}\circ\bm{\tau}\Big]\Psi\\
&-\frac{3}{8}a^{2}\Big[A_{2}^{2}\Big(\sum_{\lambda}\bm{n}^{(\lambda)}\circ\nabla^{2}\bm{n}^{(\lambda)}\Big)\tau_{0}+A_{1}A_{2}\Big(\sum_{\lambda}\nabla^{2}\bm{n}^{(\lambda)}\Big)\circ\bm{\tau}+iA_{2}^{2}\Big(\sum_{\lambda}\bm{n}^{(\lambda)}\bm{\times}\nabla^{2}\bm{n}^{(\lambda)}\Big)\circ\bm{\tau}\Big]\Psi\nonumber\\
&+\frac{3}{4}a^{2}(A_{2}/S)^{2}\Big[\big(\bm{m}\circ\partial_{x}\bm{m}\big)\tau_{0}\partial_{x}\Psi+\big(\bm{m}\circ\partial_{y}\bm{m}\big)\tau_{0}\partial_{y}\Psi\Big]\nonumber\\
&+\frac{3}{2}a^{2}(A_{1}A_{2}/S)\Big[\big(\partial_{x}\bm{m}\circ\bm{\tau}\big)\partial_{x}\Psi+\big(\partial_{y}\bm{m}\circ\bm{\tau}\big)\partial_{y}\Psi\Big]\nonumber\\
&-\frac{3}{4}ia^{2}A_{2}^{2}\Big[\Big(\sum_{\lambda}\bm{n}^{(\lambda)}\bm{\times}\partial_{x}\bm{n}^{(\lambda)}\Big)\circ\bm{\tau}\partial_{x}\Psi+\Big(\sum_{\lambda}\bm{n}^{(\lambda)}\bm{\times}\partial_{y}\bm{n}^{(\lambda)}\Big)\circ\bm{\tau}\partial_{y}\Psi\Big]\nonumber\\
&+\frac{3}{4}ia^{2}(A_{2}/S)^{2}\Big[\big(\bm{m}\bm{\times}\partial_{x}\bm{m}\big)\circ\bm{\tau}\partial_{x}\Psi+\big(\bm{m}\bm{\times}\partial_{y}\bm{m}\big)\circ\bm{\tau}\partial_{y}\Psi\Big]\nonumber\\
&+\frac{3}{8}a^{2}\Big[3(2A_{1}^{2}-A_{2}^{2})+(A_{2}/S)^{2}\bm{m}^{2}+4(A_{1}A_{2}/S)\bm{m}\circ\bm{\tau}\Big]\nabla^{2}\Psi\Bigg\}(\vec{r}_{k})+\textrm{h.c.}\nonumber
\end{align}
\end{widetext}
Here, we have assumed that $t_{ij}\equiv -t$ between NN, with $t>0$. Furthermore, we have also approximated the values of the spin and Fermi fields at the vertices of each plaquette by their value at its center, since we are interested in the long-wavelength limit. Therefore, summation over sublattices has been recasted as summation over the magnetic lattice $\Delta_{c}$, see Fig.~\ref{Fig2}, according to the prescription $\sum_{\lambda=1}^{3}\sum_{i\in\Delta_{\lambda}}\simeq\sum_{k\in\Delta_{c}}\sum_{\lambda=1}^{3}$.

Next, we will work out the terms $\sum_{\lambda}\bm{n}^{(\lambda)}\bm{\times}\partial_{\zeta}\bm{n}^{(\lambda)}$, $\sum_{\lambda}\nabla^{2}\bm{n}^{(\lambda)}$, $\sum_{\lambda}\bm{n}^{(\lambda)}\circ\nabla^{2}\bm{n}^{(\lambda)}$ and $\sum_{\lambda}\bm{n}^{(\lambda)}\bm{\times}\nabla^{2}\bm{n}^{(\lambda)}$: from Eq.~\eqref{eq:spin_expansion1} we derive the expression
\begin{widetext}
\begin{align}
\label{eq:int_step4}
\sum_{\lambda=1}^{3}\bm{n}^{(\lambda)}\bm{\times}\partial_{\zeta}\bm{n}^{(\lambda)}&=\sum_{\lambda=1}^{3}R[\frak{n}_{\lambda}]\bm{\times}\partial_{\zeta}R[\frak{n}_{\lambda}]-a\sum_{\lambda=1}^{3}R[(\frak{n}_{\lambda}\circ\bm{N})\frak{n}_{\lambda}]\bm{\times}\partial_{\zeta}R[\frak{n}_{\lambda}]\\
&\hspace{1cm}-a\sum_{\lambda=1}^{3}R[\frak{n}_{\lambda}]\bm{\times}\partial_{\zeta}R[(\frak{n}_{\lambda}\circ\bm{N})\frak{n}_{\lambda}]+a^{2}\sum_{\lambda=1}^{3}R[(\frak{n}_{\lambda}\circ\bm{N})\frak{n}_{\lambda}]\bm{\times}\partial_{\zeta}R[(\frak{n}_{\lambda}\circ\bm{N})\frak{n}_{\lambda}],\nonumber\\
&=3\bm{\Omega}_{\zeta}+\frac{3ia}{2}\Big[N_{x}\big(\bm{\Sigma}_{xy}^{\zeta}+\bm{\Sigma}_{yx}^{\zeta}\big)+N_{y}\big(\bm{\Sigma}_{xx}^{\zeta}-\bm{\Sigma}_{yy}^{\zeta}\big)\Big]\nonumber\\
&\hspace{1cm}+\frac{3ia^{2}}{8}\Big[N_{x}^{2}\big(3\bm{\Sigma}_{xx}^{\zeta}+\bm{\Sigma}_{yy}^{\zeta}\big)+2N_{x}N_{y}\big(\bm{\Sigma}_{xy}^{\zeta}+\bm{\Sigma}_{yx}^{\zeta}\big)+N_{y}^{2}\big(3\bm{\Sigma}_{yy}^{\zeta}+\bm{\Sigma}_{xx}^{\zeta}\big)\Big],\nonumber
\end{align}
\end{widetext}
where we have exploited the identities $\sum_{\lambda}\frak{n}_{\lambda}=\bm{0}$ and $\sum_{\lambda}(\frak{n}_{\lambda}\circ\bm{A})\frak{n}_{\lambda}=\frac{3}{2}\bm{A}$. Here, $\bm{\Omega}_{\zeta}=\frac{i}{2}\textrm{Tr}\left[\mathcal{P}\big(R^{\top}\bm{L}\partial_{\zeta}R\big)\right]$ represent the Yang-Mills fields for the triangular antiferromagnet, with $\mathcal{P}=\textrm{diag}(1,1,0)$ being a projector operator onto the $xy$ plane, and $\bm{\Sigma}^{\zeta}$ denotes the vector of matrix products $R^{\top}\bm{L}\partial_{\zeta}R$. Similarly, we also obtain the expressions
\begin{widetext}
\begin{align}
\sum_{\lambda=1}^{3}\nabla^{2}\bm{n}^{(\lambda)}&=\sum_{\lambda=1}^{3}\nabla^{2}R[\frak{n}_{\lambda}]+\nabla^{2}R\left[3\bm{N}-\sum_{\lambda=1}^{3}(\frak{n}_{\lambda}\circ\bm{N})\frak{n}_{\lambda}\right]+\partial_{\zeta}R\left[3\partial_{\zeta}\bm{N}-\sum_{\lambda=1}^{3}(\frak{n}_{\lambda}\circ\partial_{\zeta}\bm{N})\frak{n}_{\lambda}\right]\\
&\hspace{2cm}+R\left[3\nabla^{2}\bm{N}-\sum_{\lambda=1}^{3}(\frak{n}_{\lambda}\circ\nabla^{2}\bm{N})\frak{n}_{\lambda}\right],\nonumber\\
&=\frac{3}{2}a\left\{\nabla^{2}R[\bm{N}]+2\partial_{\zeta}R[\partial_{\zeta}\bm{N}]+R[\nabla^{2}\bm{N}]\right\}=\frac{3}{2}a\nabla^{2}\left\{R[\bm{N}]\right\},\nonumber\\
\rho\sum_{\lambda=1}^{3}\bm{n}^{(\lambda)}\circ\nabla^{2}\bm{n}^{(\lambda)}&=\partial_{\zeta}\left[\rho\sum_{\lambda=1}^{3}\bm{n}^{(\lambda)}\circ\partial_{\zeta}\bm{n}^{(\lambda)}\right]-\partial_{\zeta}\rho\sum_{\lambda=1}^{3}\bm{n}^{(\lambda)}\circ\partial_{\zeta}\bm{n}^{(\lambda)}-\rho\sum_{\lambda=1}^{3}\partial_{\zeta}\bm{n}^{(\lambda)}\circ\partial_{\zeta}\bm{n}^{(\lambda)},\\
&=-\rho\sum_{\lambda=1}^{3}\partial_{\zeta}R[\frak{n}_{\lambda}]\circ\partial_{\zeta}R[\frak{n}_{\lambda}]+O(a,|\bm{N}|)=-3\rho\,\bm{\Omega}_{\zeta}\circ\bm{\Omega}_{\zeta}+O(a,|\bm{N}|),\nonumber\\
\bm{s}\circ\sum_{\lambda=1}^{3}\bm{n}^{(\lambda)}\bm{\times}\nabla^{2}\bm{n}^{(\lambda)}&=\partial_{\zeta}\left[\bm{s}\circ\sum_{\lambda=1}^{3}\bm{n}^{(\lambda)}\bm{\times}\partial_{\zeta}\bm{n}^{(\lambda)}\right]-\partial_{\zeta}\bm{s}\circ\sum_{\lambda=1}^{3}\bm{n}^{(\lambda)}\bm{\times}\partial_{\zeta}\bm{n}^{(\lambda)}-\bm{s}\circ\sum_{\lambda=1}^{3}\partial_{\zeta}\bm{n}^{(\lambda)}\bm{\times}\partial_{\zeta}\bm{n}^{(\lambda)},\\
&=3\partial_{\zeta}\left[\bm{s}\circ\bm{\Omega}_{\zeta}\right]-3\bm{\Omega}_{\zeta}\circ\partial_{\zeta}\bm{s}+O(a,|\bm{N}|)=3(\partial_{\zeta}\bm{\Omega}_{\zeta})\circ\bm{s}+O(a,|\bm{N}|),\nonumber
\end{align}
\end{widetext}
where we have also taken into account the identities $\bm{n}^{(\lambda)}\circ\partial_{\zeta}\bm{n}^{(\lambda)}=\partial_{\zeta}[\bm{n}^{(\lambda)}]^{2}/2=0$ and $\textrm{Tr}\big[\partial_{\zeta}R^{\top}\partial_{\zeta}R\big]/2=\bm{\Omega}_{\zeta}\circ\bm{\Omega}_{\zeta}$. Consequently, we finally obtain the following mean-field expression for the tight-binding Hamiltonian in the SRI SB representation:
\begin{widetext}
\begin{align}
\label{eq:tb_hamiltonian_MF}
\langle H_{\textrm{tb}}^{\textrm{SRI SB}}\rangle_{\textrm{MF}}=&-t\sum_{k\in\Delta_{c}}\Psi^{\dagger}\Bigg\{\bigg[\frac{3}{2}\big[3(2A_{1}^{2}-A_{2}^{2})+(A_{2}/S)^{2}\bm{m}^{2}\big]\tau_{0}+6(A_{1}A_{2}/S)\bm{m}\circ\bm{\tau}\bigg]\Psi\\
&+\frac{9}{8}a^{2}A_{2}^{2}\bigg[\vec{\bm{\Omega}}\bm{\odot}\vec{\bm{\Omega}}\,\tau_{0}-i\partial_{\mu}\bm{\Omega}_{\mu}\circ\bm{\tau}\bigg]\Psi-\frac{9}{4}ia^{2}A_{2}^{2}\Big[\big(\bm{\Omega}_{x}\circ\bm{\tau}\big)\partial_{x}\Psi+\big(\bm{\Omega}_{y}\circ\bm{\tau}\big)\partial_{y}\Psi\Big]\nonumber\\
&+\frac{3}{8}a^{2}\Big[3(2A_{1}^{2}-A_{2}^{2})+(A_{2}/S)^{2}\bm{m}^{2}+4(A_{1}A_{2}/S)\bm{m}\circ\bm{\tau}\Big]\nabla^{2}\Psi\Bigg\}(\vec{r}_{k})+\textrm{h.c.}+O(|\bm{m}|,\partial)^{3},
\end{align}
\end{widetext}
where we are disregarding higher order terms than quadratic in the magnetization and partial derivatives.

Within the path-integral approach the kinetic term of the Lagrangian reads $\hbar\sum_{i,\sigma}c_{i,\sigma}^{\dagger}\partial_{\tau}c_{i,\sigma}$. By incorporating the SRI SB representation for electron operators into it, we obtain at the mean-field level
\begin{align}
\Big\langle\sum_{i,\sigma}c_{i,\sigma}^{\dagger}\partial_{\tau}c_{i,\sigma}\Big\rangle_{\textrm{MF}}&=\sum_{i}\Psi_{i}^{\dagger}\big[\langle\underline{z_{i}}\rangle_{\textrm{MF}}^{\dagger}\partial_{\tau}\langle\underline{z_{i}}\rangle_{\textrm{MF}}\big]\Psi_{i}\\
&\hspace{0.5cm}+\sum_{i}\Psi_{i}^{\dagger}\big[\langle\underline{z_{i}}\rangle_{\textrm{MF}}^{\dagger}\langle\underline{z_{i}}\rangle_{\textrm{MF}}\big]\partial_{\tau}\Psi_{i}.\nonumber
\end{align}
With account of Eq.~\eqref{eq:matrix_z_MF3}, the mean-field value of the above matrix products becomes
\begin{align}
\label{eq:rel5}
\langle\underline{z_{i}}\rangle_{\textrm{MF}}^{\dagger}\langle\underline{z_{i}}\rangle_{\textrm{MF}}&=\big(A_{1}^{2}+A_{2}^{2}\big)\tau_{0}+2A_{1}A_{2}\bm{n}(\vec{r}_{i})\circ\bm{\tau},\\
\label{eq:rel6}
\langle\underline{z_{i}}\rangle_{\textrm{MF}}^{\dagger}\partial_{\tau}\langle\underline{z_{i}}\rangle_{\textrm{MF}}&=\Big[A_{1}A_{2}\partial_{\tau}\bm{n}+iA_{2}^{2}\bm{n}\bm{\times}\partial_{\tau}\bm{n}\Big](\vec{r}_{i})\circ\bm{\tau},
\end{align}
where in the last equation we have used that $\bm{n}\circ\partial_{\tau}\bm{n}=0$ due to the normalization of the spin field. Therefore, we obtain
\begin{widetext}
\begin{align}
\label{eq:kin_term2}
\sum_{i}\Psi_{i}^{\dagger}\big[\langle\underline{z_{i}}\rangle_{\textrm{MF}}^{\dagger}\langle\underline{z_{i}}\rangle_{\textrm{MF}}\big]\partial_{\tau}\Psi_{i}&=\sum_{\lambda=1}^{3}\sum_{i\in\Delta_{\lambda}}\Psi_{i}^{\dagger}\big[\big(A_{1}^{2}+A_{2}^{2}\big)\tau_{0}+2A_{1}A_{2}\bm{n}^{(\lambda)}(\vec{r}_{i})\circ\bm{\tau}\big]\partial_{\tau}\Psi_{i}\simeq3(A_{1}^{2}+A_{2}^{2})\sum_{k\in\Delta_{c}}\big[\Psi^{\dagger}\partial_{\tau}\Psi\big](\vec{r}_{k}).
\end{align}
\end{widetext}
and
\begin{widetext}
\begin{align}
\label{eq:kin_term}
\sum_{i}\Psi_{i}^{\dagger}\big[\langle\underline{z_{i}}\rangle_{\textrm{MF}}^{\dagger}&\partial_{\tau}\langle\underline{z_{i}}\rangle_{\textrm{MF}}\big]\Psi_{i}\simeq\sum_{k\in\Delta_{c}}\sum_{\lambda=1}^{3}\Psi^{\dagger}(\vec{r}_{k})\Big[A_{1}A_{2}\partial_{\tau}\bm{n}^{(\lambda)}+iA_{2}^{2}\,\bm{n}^{(\lambda)}\bm{\times}\partial_{\tau}\bm{n}^{(\lambda)}\Big](\vec{r}_{k})\circ\bm{\tau}\Psi(\vec{r}_{k}),\\
&\simeq\sum_{k\in\Delta_{c}}\bigg\{3iA_{2}^{2}\,\bm{\Omega}_{\tau}\circ\big(\Psi^{\dagger}\bm{\tau}\Psi\big)-\frac{3a}{2}A_{2}^{2}\Big[N_{x}\big(\bm{\Sigma}_{xy}^{\tau}+\bm{\Sigma}_{yx}^{\tau}\big)+N_{y}\big(\bm{\Sigma}_{xx}^{\tau}-\bm{\Sigma}_{yy}^{\tau}\big)\Big]\circ\big(\Psi^{\dagger}\bm{\tau}\Psi\big)\nonumber\\
&\hspace{0.5cm}-\frac{3a^{2}}{8}A_{2}^{2}\Big[N_{x}^{2}\big(3\bm{\Sigma}_{xx}^{\tau}+\bm{\Sigma}_{yy}^{\tau}\big)+2N_{x}N_{y}\big(\bm{\Sigma}_{xy}^{\tau}+\bm{\Sigma}_{yx}^{\tau}\big)+N_{y}^{2}\big(3\bm{\Sigma}_{yy}^{\tau}+\bm{\Sigma}_{xx}^{\tau}\big)\Big]\circ\big(\Psi^{\dagger}\bm{\tau}\Psi\big)\bigg\}(\vec{r}_{k}),\nonumber\\
&\simeq\sum_{k\in\Delta_{c}}\big[3iA_{2}^{2}\,\bm{\Omega}_{\tau}\circ\big(\Psi^{\dagger}\bm{\tau}\Psi\big)\big](\vec{r}_{k})+O(|\bm{m}|,\partial)^{2}.\nonumber
\end{align}
\end{widetext}
Here, we have disregarded those terms in the kinetic Lagrangian that are quadratic (or higher order) in the magnetization and partial derivatives. Finally, the continuum limit of the low-energy long-wavelength theory is achieved by means of the Riemann's prescription $\sum_{k\in\Delta_{c}}\simeq\frac{1}{S(\Delta_{c})}\int dxdy$, where $S(\Delta_{c})=3\sqrt{3}a^{2}/2$ is the area of a magnetic unit cell. The corresponding kinetic Lagrangian and tight-binding Hamiltonian then read
\begin{widetext}
\begin{align}
\label{eq:kin_term_cont}
\langle L_{\textrm{kin}}[t]\rangle_{\textrm{MF}}&=\frac{2}{3\sqrt{3}a^{2}}\int dxdy\hspace{0.1cm}\Psi^{\dagger}\big[3(A_{1}^{2}+A_{2}^{2})\hbar\partial_{\tau}+i\hbar\big(3A_{2}^{2}\,\bm{\Omega}_{\tau}\big)\circ\bm{\tau}\big]\Psi+O(|\bm{m}|,\partial)^{2},\\
\langle H_{\textrm{tb}}^{\textrm{SRI SB}}\rangle_{\textrm{MF}}&=-\frac{2}{3\sqrt{3}a^{2}}t\int dxdy\hspace{0.1cm}\bigg[9(2A_{1}^{2}-A_{2}^{2})|\Psi|^{2}+3(A_{2}/S)^{2}\bm{m}^{2}|\Psi|^{2}+12(A_{1}A_{2}/S)\bm{m}\circ\Psi^{\dagger}\bm{\tau}\Psi+\frac{9}{4}a^{2}A_{2}^{2}\,\vec{\bm{\Omega}}\bm{\odot}\vec{\bm{\Omega}\,|\Psi|^{2}}\nonumber\\
&\hspace{0.5cm}+\frac{9}{4}a^{2}A_{2}^{2}\,\bm{\Omega}_{\mu}\circ\big[i\partial_{\mu}\Psi^{\dagger}\bm{\tau}\Psi-i\Psi^{\dagger}\bm{\tau}\partial_{\mu}\Psi\big]+\frac{9}{8}(2A_{1}^{2}-A_{2}^{2})a^{2}(\Psi^{\dagger}\nabla^{2}\Psi+\nabla^{2}\Psi^{\dagger}\Psi)\bigg]+O(|\bm{m}|,\partial)^{3},
\end{align}
\end{widetext}
We introduce next the dimensionless coordinates $\vec{r}\rightarrow\vec{r}/\ell$, $\ell=\sqrt[4]{3}a/\sqrt{2(A_{1}^{2}+A_{2}^{2})}$, so that the above expansions become
\begin{widetext}
\begin{align}
\label{eq:kin_term_cont2}
\langle L_{\textrm{kin}}[t]\rangle_{\textrm{MF}}&=\int dxdy\hspace{0.1cm}\Psi^{\dagger}\big[\hbar\partial_{\tau}+ig_{0}\bm{\Omega}_{\tau}\circ\bm{\tau}\big]\Psi+O(|\bm{m}|,\partial)^{2},\\
\label{eq:Ham_cont2}
\langle H_{\textrm{tb}}^{\textrm{SRI SB}}\rangle_{\textrm{MF}}&=\int dxdy\hspace{0.1cm}\bigg\{\frac{\hbar^{2}}{2m_{\star}}\nabla\Psi^{\dagger}\cdot\nabla\Psi+g_{1}\bm{\Omega}_{\mu}\circ\big[i\Psi^{\dagger}\bm{\tau}\partial_{\mu}\Psi-i\partial_{\mu}\Psi^{\dagger}\bm{\tau}\Psi\big]-\big(g_{2}'+g_{2}\bm{m}^{2}+g_{3}\vec{\bm{\Omega}}\bm{\odot}\vec{\bm{\Omega}}\big)|\Psi|^{2}\\
&\hspace{2.0cm}-g_{4}\bm{m}\circ\Psi^{\dagger}\bm{\tau}\Psi\bigg\}+O(|\bm{m}|,\partial)^{3},\nonumber
\end{align}
\end{widetext}
with the coupling constants $g$'s and effective mass $m_{\star}$ reading
\begin{widetext}
\begin{align}
\label{eq:couplings}
&g_{0}=\frac{A_{2}^{2}}{A_{1}^{2}+A_{2}^{2}}\hbar,\hspace{0.2cm}\frac{\hbar^{2}}{2m_{\star}}=\frac{\sqrt{3}}{2}(2A_{1}^{2}-A_{2}^{2})t,\hspace{0.2cm}g_{1}=\frac{\sqrt{3}}{2}A_{2}^{2}t,\hspace{0.2cm}g_{2}'=3\frac{2A_{1}^{2}-A_{2}^{2}}{A_{1}^{2}+A_{2}^{2}}t,\\
&g_{2}=\frac{A_{2}^{2}}{A_{1}^{2}+A_{2}^{2}}\frac{t}{S^{2}},\hspace{0.2cm}g_{3}=\frac{\sqrt{3}}{2}A_{2}^{2}t\hspace{0.2cm}g_{4}=\frac{A_{1}A_{2}}{A_{1}^{2}+A_{2}^{2}}\frac{4t}{S}.\nonumber
\end{align}
\end{widetext}
We note in passing that the term $g_{2}'|\Psi|^{2}$ in the above effective Hamiltonian can be disregarded since it simply contributes with an energy shift to the total energy functional ($\int d\vec{r}\,|\Psi|^{2}=1$ due to the normalization condition of the wave function).

\section{Discussion}
\label{Disc}

In this work we have constructed a transport theory for itinerant carriers flowing within a magnetically frustrated conductor and studied the spin-transfer physics resulting from the exchange of angular momentum between the localized and itinerant spin degrees of freedom. In this regard, it is worth remarking that our theory also incorporates the description of charge and spin transport in conventional antiferromagnets: there is a (canonical) embedding of the N\'{e}el order, whose manifold is the three-dimensional unit sphere, into the group of unit-norm quaternions, given by $\bm{l}\mapsto\mathbf{q}_{l}=(0,\bm{l})$. The corresponding Yang-Mills fields read $\bm{\Omega}_{\mu}=2\partial_{\mu}\mathbf{q}_{l}\wedge\mathbf{q}_{l}^{\star}=2\bm{l}\bm{\times}\partial_{\mu}\bm{l}$, i.e., they are proportional to the antiferromagnetic magnetization currents. By substituting these expressions into Eq.~\eqref{eq:Ham_cont2}, we retrieve the effective Hamiltonian for mobile vacancies in the two-dimensional Heisenberg antiferromagnet derived by Shraiman and Siggia in the context of high-$T_{\textrm{c}}$ superconductivity \cite{Shraiman-PRL1988}. Furthermore, the resultant expression for the magnetic torque~\eqref{eq:mag_torque1} reads $\partial_{t}\bm{m}\propto j^{e}_{\mu}\bm{l}\bm{\times}\partial_{\mu}\bm{l}=\bm{l}\bm{\times}\big(\vec{j}^{e}\cdot\nabla\big)\bm{l}$, which coincides with that for the (dissipative) spin-transfer torque acting on a bipartite antiferromagnet \cite{Baltz-RMP2018}. We mention that a low-energy long-wavelength theory for itinerant carriers in a (bipartite) antiferromagnet, whose effective Hamiltonian is given by that of Ref.~\onlinecite{Shraiman-PRL1988}, can be built along the lines of the present work, also based upon the SB representation, if one accounts for the Haldane mapping $\bm{n}(\vec{r}_{i})=\lambda_{i}\,\bm{l}(\vec{r}_{i})\sqrt{1-\bm{m}^{2}(\vec{r}_{i})}+\bm{m}(\vec{r}_{i})$ instead of Eq.~\eqref{eq:spins} for the unit-norm spin field ($\lambda_{i}=\pm1$ on different sublattices). This illustrates how powerful and well-suited the SB technique is to describe the spin-transfer physics in a generic magnet, since the interplay between localized and itinerant spin degrees of freedom is naturally incorporated in the representation of the electron operators.

We note in passing that many effective low-energy theories for charge carriers (spin$-\tfrac{1}{2}$ fermions) in solid-state systems can be cast into the Yang-Mills form. One important example is that describing the dynamics of electrons in a homogeneous medium with weak spin-orbit coupling \cite{Jin-JPA2006}, which provided one of the first unified descriptions of charge and spin transport in spin-orbit coupled systems at a time of development of the spin Hall effect. Therefore, it is worth mentioning the key differences between our effective Yang-Mills theory for magnetically frustrated conductors and the aforesaid one. The physical systems described by each theory are entirely distinct and different in physical origin (magnetic frustration vs. spin-orbit interactions). Indeed, the nature of the corresponding Yang-Mills fields in each of the theories are physically unrelated, since in Ref.~\onlinecite{Jin-JPA2006} the gauge fields are rooted in the weak spin-orbit interaction, whereas in our work the Yang-Mills fields arise from exchange coupling of the itinerant spin degrees of freedom to the noncollinear magnetic background. 
 In addition, the spin torques discussed in our work act on the macroscopic spin density associated with the localized spin degrees of freedom, whereas in Ref.~\onlinecite{Jin-JPA2006} forces/torques act on the itinerant charge and spin densities.

 \begin{figure}[h!]
\begin{center}
\includegraphics[width=1.0\columnwidth]{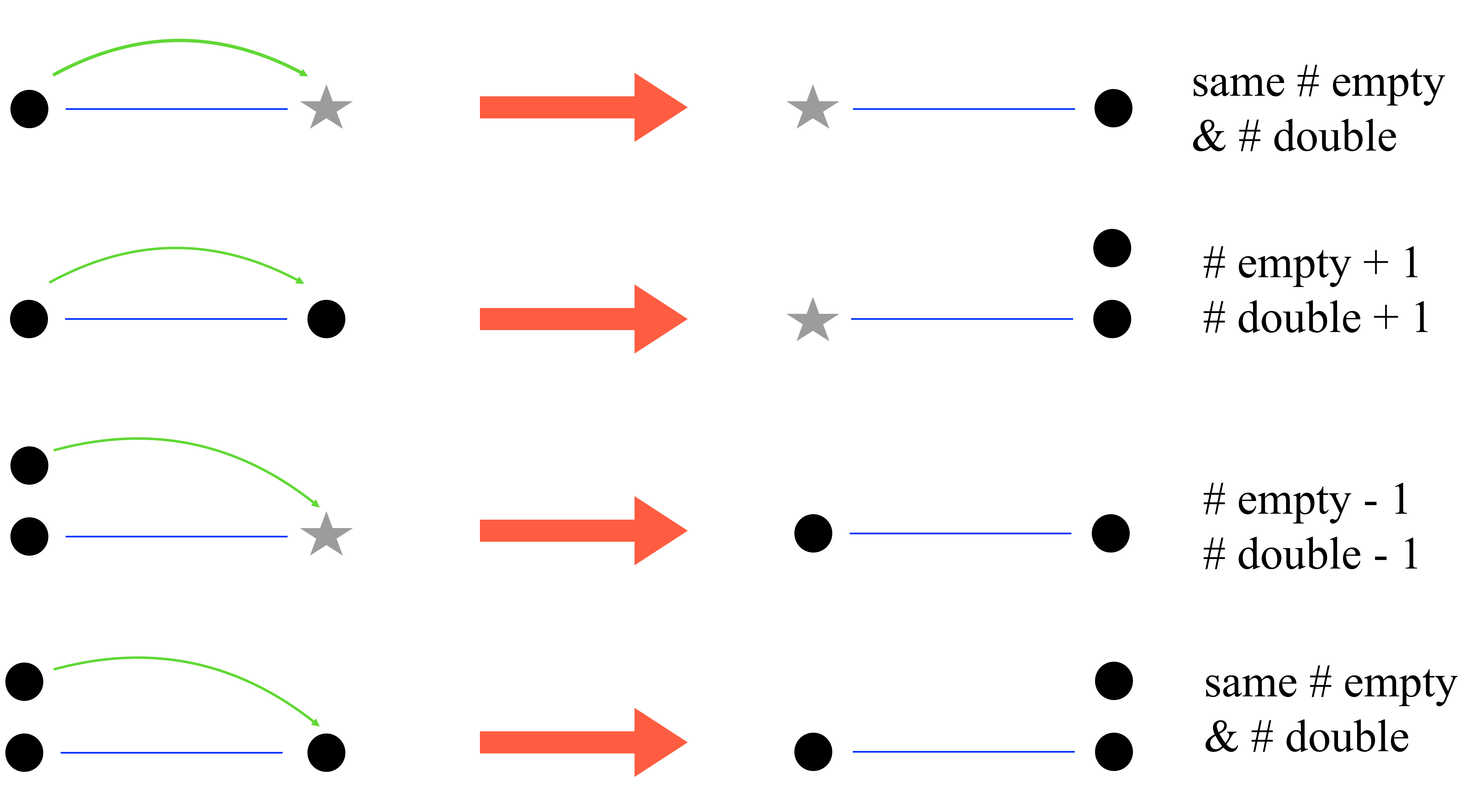}
\caption{Electron hopping processes in the half-filled limit. Circles and stars denote electrons and holes, respectively.}
\vspace{-0.5cm} 
\label{Fig3}
\end{center}
\end{figure}

\subsection*{Half-filled limit}

For the sake of correctness, it is interesting to elucidate the half-filled limit of our effective theory, described by the Lagrangian~\eqref{eq:Lag}. We note first that the Hubbard model presents one electron per lattice site (on average) at half filling. A simple analysis of (electron) hopping processes in this limit, see Fig. \ref{Fig3}, points to the conclusion that the number of empty sites equals that of double occupancies, i.e, $e=d$ always holds at the mean-field level in this limit. As a result, the auxiliary functions $a(\sigma_{1},\sigma_{2})$ become independent of the parameter $\sigma_{1}$, $a(\sigma_{1},\sigma_{2})=1/\sqrt{1+2\sigma_{2}s_{0}S}$, which leads the quadratic combinations
\begin{equation}
a(+,-)a(-,+)\pm a(+,+)a(-,-)=\left\{
    \begin{array}{l}
     \frac{2}{\sqrt{1-4s_{0}^{2}S^{2}}},\\
      0,
    \end{array} \right.
\end{equation}
to simplify enormously. The coefficients defined in Eq.~\eqref{eq:matrix_z_MF3} now become $A_{1}(s_{0},d,d)=2\sqrt{2}s_{0}d/\sqrt{1-4s_{0}^{2}S^{2}}$ and $A_{2}(s_{0},d,d)=0$, so that the coupling constants $g_{0},g_{1},g_{2},g_{3}$ and $g_{4}$ are identically zero, see Eq.~\eqref{eq:couplings}. We therefore conclude that the effective Lagrangian for itinerant carriers reduces, in the half-filled limit, to
\begin{equation}
\mathcal{L}=\Psi^{\dagger}\hbar\partial_{\tau}\Psi+\frac{\hbar^{2}}{2m_{\star}}\nabla\Psi^{\dagger}\nabla\Psi-g_{2}'|\Psi|^{2},
\end{equation}
with
\begin{equation}
\frac{\hbar^{2}}{2m_{\star}}=\frac{8\sqrt{3}s_{0}^{2}d^{2}}{1-4s_{0}^{2}S^{2}}t,\textrm{ and }g_{2}'=6t.
\end{equation}
In particular, the energy functional of the system reads
\begin{equation}
\label{eq:E_half_filling}
\mathcal{E}[\Psi]=\int d\vec{r}\left[\frac{\hbar^{2}}{2m_{\star}}\nabla\Psi^{\dagger}\nabla\Psi-g_{2}'|\Psi|^{2}\right].
\end{equation}
A time-independent plane-wave description of the pseudofermions, parametrized by the crystal momentum $\vec{k}$, yields the following expression for the energy density:
\begin{equation}
\epsilon[\vec{k}\hspace{0.05cm}]=\frac{\hbar^{2}k^{2}}{2m_{\star}}-g_{2}',
\end{equation}
which corresponds to the energy of a free (massive) particle, with the effective mass being inversely proportional to the NN hopping integral. We note that the absence of other energy terms in Eq. \eqref{eq:E_half_filling} depending on the magnetization and/or the Yang-Mills fields is rooted in the fact that, in the half-filled limit, the mean-field value of the matrix operator $\underline{z_{i}}$, see Eq. \eqref{eq:matrix_z_MF3}, is proportional to the identity matrix in the spin space and also independent of the background spin field. As a final remark, when the limit of large exchange coupling is also considered, electrons are polarized along the direction of the localized magnetic moments (namely, the noncollinear magnetic background), which, in turn, leads to the absence of double electronic occupancies in the ground state as one of the spin polarizations is always energetically penalized. Therefore, $e=d=0$ and the itinerant charge fluid exhibits a dispersionless (flat) energy band in this limit; in other words, the frustrated magnet becomes an insulator.

\subsection*{External magnetic field}

We can account in our framework for the effects of an external magnetic field $\bm{B}$ as follows: to begin with, for moderate field strengths (moderate in the sense that the Zeeman term competes with the geometric frustration but does not dominate energetically), we can incorporate the magnetic field into the description of the magnetically frustrated background via the substitution $\partial_{t}R\mapsto\partial_{t}R-i\gamma(\bm{B}\circ\bm{L})R$, where $\gamma$ denotes the gyromagnetic ratio \cite{Ochoa-PRB2018}. In our transport theory, it only affects the angular velocity $\bm{\Omega}_{t}$ of the localized spin subsystem, i.e., $\bm{\Omega}_{t}\mapsto\bm{\Omega}_{t}+\gamma\bm{B}$. Therefore, the angular velocity of the magnetic background is shifted by the Larmor frequency, as expected. We note that this shift turns into a Zeeman-like term in the Lagrangian~\eqref{eq:Lag} for itinerant carriers. Second, the effect of the magnetic field on the itinerant subsystem can be captured by means of the conventional minimal coupling $\vec{p}\mapsto\vec{p}-e\vec{A}$, where $\vec{A}$ denotes the corresponding electromagnetic gauge field. Third, reciprocity arguments can still be applied if we take into account the Onsager-Casimir relations $\hat{L}_{\textrm{sq}}(\bm{m},\bm{B})=-\hat{L}_{\textrm{qs}}^{\top}(-\bm{m},-\bm{B})$ for the Onsager coefficients \cite{Mazur-Physica1953}. We note that the presence of a magnetic field does not affect our expressions for the spin-transfer torques. Furthermore, we have disregarded any magnetic torque coupling the magnetic field to the Yang-Mills fields since we are interested in the spin-transfer effects emerging only from the noncollinear magnetic background.

\subsection*{Role of the Hubbard $U$ term}

It is important to mention that we have not dismissed the on-site Coulomb repulsion term in the transport theory we are presenting here. To begin with, we note that, regardless of its microscopic origin, the effective low-energy long-wavelength Lagrangian for itinerant carriers, Eq.~\eqref{eq:Lag}, can also be derived phenomenologically (namely, based solely on symmetry grounds). In our work we have shown that the kinetic term of the Hubbard model, which describes the electron motion within the lattice, combined with the slave-boson approach accounting for the backflow of spin excitations during the hopping process, is enough to reproduce the aforementioned Lagrangian in the long-wavelength limit. Furthermore, the Hubbard $U$ term  becomes quartic in the Fermi field (in the SB representation), which may be responsible for instabilities and possible pairing of the itinerant carriers. The resultant physics go beyond the scope of this work, and therefore the Coulomb term only contributes here to determining which is the ground-state phase of the system (and, in turn, to the mean-field values $e$, $d$ of the bosonic occupancies).

\subsection*{Role of fluctuations}

In the presence of a continuous (spin rotation) symmetry, fluctuations preclude the onset of a long-range (magnetic) order in one- and two-dimensional systems at nonzero temperature according to the Mermin-Wagner-Hohenberg-Coleman theorem. We note that this does not represent a caveat for our transport theory: as discussed in Sec.~\ref{Frust_mag}, the spin rotation symmetry of frustrated magnets emerges at mesoscopic length scales as a result of the Kadanoff-Wilson-type renormalization procedure. This symmetry is broken microscopically by the presence of magnetocrystalline anisotropies, an external magnetic field or, more generally, other (spin) rotation symmetry-breaking interactions, which open a gap in the excitation spectrum and lift some of the Goldstone modes \cite{Benefatto-PRB2006,Tserkovnyak-PRB2017,Ochoa-PRB2018}, yielding for instance the dynamical stabilization of the magnetic order\cite{Zarzuela-PRB2021}. Furthermore, we mention that the SO(3) group of rotations (describing the optimal mapping between two quasi-degenerate ground states) has a nontrivial first homotopy group, $\pi_{1}(\textrm{SO}(3))=\mathbb{Z}_{2}$. This means that the order-parameter manifold of frustrated magnets hosts one-dimensional topological defects, the so-called magnetic disclinations, which play a similar role to that of magnetic vortices in the $XY$-model for collinear magnets. Therefore, if the spin rotation symmetry was exact, we would expect the existence of a quasi-long-range ordered phase below a certain critical temperature in the localized spin sector of a frustrated magnet.

In this work, we have addressed electronic correlations via the mean-field picture. Our justification is based on the fact that variations of the order-parameter of the localized spin subsystem occur on mesoscopic length scales, so that itinerant carriers 'see' a uniform magnetic background microscopically. We therefore consider that the slave-bosonic occupation numbers should also be averaged out at mesoscopic length scales in our effective theory: fluctuations of the magnetic background will affect our derivations at length scales larger than mesoscopic ones, since these fluctuations correspond to those of the order-parameter rotation matrix. By assuming the smoothness of the slave-bosonic occupation numbers at mesoscopic length scales, from our microscopic derivation we can conclude that fluctuations lead to a spatial dependence of the coupling constants $\{m_{\star}(\vec{r}),g_{0}(\vec{r}),\ldots,g_{4}(\vec{r})\}$, in the spirit of the renormalization-group approach.

\subsection*{Hall physics and experimental probes}

The term proportional to $g_{1}$ in the effective Lagrangian~\eqref{eq:Lag}, which couples the (background) magnetization currents to the spin current carried by the itinerant electrons, also determines the topological Hall response of frustrated magnets. Indeed, since our low-energy long-wavelength theory is of the Yang-Mills type, a (nonabelian) effective electromagnetic field emerges in the magnet, which affects the dynamics of itinerant carriers akin to the case of conventional magnetic conductors. That is to say, both charge and spin Hall currents will be created when charge flows within the frustrated magnet. In the case of conventional magnets ($S^{2}$-order parameter), the topological Hall effect has an orbital origin, since it can be understood as resulting from the action of a (topological) magnetic field on spinless carriers (electrons follow the spin texture adiabatically). In these conventional magnets, the nondiagonal components of the effective gauge field tensor can be disregarded due to the large exchange coupling, giving rise to the orbital character of the topological Hall effect. In the case of frustrated magnets, however, the spin of the carriers do not follow the background spin texture adiabatically (as spin fluctuations usually happen in short length scales) and the adiabaticity is only recovered at the mesoscopic level (after coarse-graining). 

We argue that nitrogen-vacancy center magnetometry can probe the spin chirality in these systems, over length scales corresponding to the size of the glassy domains, as well as the emergent topological solitons (namely, Shankar skyrmions\cite{Shankar-JPhys1977,Volovik-JETP1977} and $4\pi$-vortices\cite{Anderson-PRL1976}) and defects (magnetic disclinations). These topological SO(3) textures mediate a topological Hall response in magnetically frustrated platforms, engendering both charge and spin Hall currents that can be detected experimentally by standard means/Hall geometry. Similarly, they also mediate a new contribution to the spin-transfer torque acting on the macroscopic spin density of the magnet. Remarkably, this magnetic torque, along with the disclination-mediated Hall currents, have no counterparts in collinear magnetism, representing therefore a novel hallmark of topological transport in frustrated spin systems. Further discussion of the spin-transfer and topological Hall responses in frustrated magnets can be found in a companion paper \cite{Zarzuela-PRL2021}.

\section{Acknowledgements}

This work has been supported by the Transregional Collaborative Research Center (SFB/TRR) 173 SPIN+X, the skyrmionics SPP: ZA1194/2-1, SI1720/12-1, Grant Agency of the Czech Republic grant no. 19-28375X, and the Dynamics and Topology Centre funded by the State of Rhineland Palatinate. This research was supported in part by the National Science Foundation under Grant No. NSF PHY-1748958.
 
\appendix

\section{Spin-rotation invariant slave boson representation}
\label{AppA}

The slave boson (SB) representation can be formally realized by embedding, at each lattice site, the physical Hilbert space of (electronic) fermion states into a larger Hilbert space containing auxiliary bosons that account for the aforementioned spin backflow when electrons hop. The original SB formalism introduced by Kotliar and Ruckenstein \cite{Kotliar-PRL1986} explicitly breaks the spin rotational invariance of the Hubbard model (due to the assumption of a spin quantization axis in the boson space) and, therefore, may be misleading when constructing any mean-field theory involving magnetic textures (i.e., for spatially nonuniform spin quantization axes). 

In the spin-rotation invariant (SRI) extension proposed in Ref.~\onlinecite{Li-PRB1989}, the physical states are obtained by creating electrons and auxiliary bosons in the vacuum:
\begin{align}
\label{eq:A1}
|i,0\rangle&=e^{\dagger}_{i}|0\rangle,\\
|i,\sigma\rangle&=s^{\dagger}_{i,\sigma}f^{\dagger}_{i,\sigma}|0\rangle,\nonumber\\
|i,\uparrow\downarrow\rangle&=d^{\dagger}_{i}f^{\dagger}_{i,\uparrow}f^{\dagger}_{i,\downarrow}|0\rangle,\nonumber
\end{align}
where $i$ and $\sigma=\uparrow,\downarrow$ denote the lattice site and the spin$-\frac{1}{2}$ projection onto the quantization axis, respectively. $f_{i,\sigma}$ are pseudofermion operators and the four boson operators $\{e_{i},s_{i,\uparrow},s_{i\downarrow},d_{i}\}$ label all possible occupations at the $i$-th lattice site, namely, \emph{empty}, \emph{single} occupation with spin $\uparrow,\downarrow$ and \emph{double} occupation, respectively. Under rotations $\,\mathcal{U}$ in the spin space, $\{|i,0\rangle, |i,\uparrow\downarrow\rangle\}$ must transform as scalars, while the single occupied states are required to transform as spinor states:
\begin{equation}
\label{eq:A2}
|i,\sigma\rangle\rightarrow\sum_{\sigma'}\mathcal{U}_{\sigma,\sigma'}^{\dagger}|i,\sigma'\rangle.
\end{equation}
By construction, the operator product $s^{\dagger}_{i,\sigma}f^{\dagger}_{i,\sigma}$ creates a composite particle of spin $1/2$; since $f_{i,\sigma}$ are pseudofermions, the rules of angular momentum coupling imply that the boson operators $s_{i,\sigma}$ must have either spin $S=0$ (the singlet component $s_{i,0}$) or spin $S=1$ (the triplet component $\bm{s}_{i}$). The couplings $\bm{0}+\bm{\frac{1}{2}}=\bm{\frac{1}{2}}$ and $\bm{1}+\bm{\frac{1}{2}}=\bm{\frac{1}{2}}$ of angular momenta are parametrized in terms of Clebsch-Gordan coefficients, which yields the following recasting of the single-occupied states:

\begin{align}
\label{eq:A4}
|i,\sigma\rangle\equiv\sum_{\sigma'=\pm\frac{1}{2}}S_{i,\sigma\sigma'}^{\dagger}f_{i\sigma'}^{\dagger}|0\rangle,
\end{align}
where the SU(2) operator $\underline{S_{i}}$ is defined as
\begin{equation}
\label{eq:A5}
\underline{S_{i}}^{\dagger}=\frac{1}{\sqrt{2}}\begin{pmatrix}
s_{i,0}^{\dagger}+s_{i,z}^{\dagger} & s_{i,x}^{\dagger}-is_{i,y}^{\dagger}\\
s_{i,x}^{\dagger}+is_{i,y}^{\dagger} & s_{i,0}^{\dagger}-s_{i,z}^{\dagger}
\end{pmatrix}=\frac{1}{\sqrt{2}}\big[s_{i,0}^{\dagger}\tau_{0}+\bm{s}_{i}\,^{\dagger}\circ\bm{\tau}\big],
\end{equation}
By projecting this expression onto the SU(2) basis $\{\tau_{0},\bm{\tau}\}$, the spin-zero and spin-one bosons can be cast as $s_{i,\lambda}^{\dagger}=\frac{1}{\sqrt{2}}\textrm{Tr}\big[\underline{S_{i}}^{\dagger}\tau_{\lambda}\big]$, $\lambda=0,x,y,z$. In this regard, it is worth remarking that the spin singlets $\{s_{i,0}\}_{i}$ and spin triplets $\{\bm{s}_{i}\}_{i}$ represent the charge and spin degrees of freedom of the spinor, respectively. Unphysical states are discarded in the SRI SB representation by imposing the local constraint
\begin{equation}
\label{eq:A6}
e_{i}^{\dagger}e_{i}+\textrm{Tr}\big[\underline{S_{i}}^{\dagger}\underline{S_{i}}\big]+d_{i}^{\dagger}d_{i}=e_{i}^{\dagger}e_{i}+s_{i,0}^{\dagger}s_{i,0}+\bm{s}_{i}\hspace{0.01cm}^{\dagger}\circ\bm{s}_{i}+d_{i}^{\dagger}d_{i}=1,
\end{equation}
on the four Bose operators, namely, only one physical state can occur per site. Furthermore, conservation of the fermion number at each lattice site yields the set of constraints
\begin{equation}
\label{eq:A7}
\textrm{Tr}\big[\tau_{\mu}\underline{S_{i}}^{\dagger}\underline{S_{i}}\big]+2\delta_{\mu0}d_{i}^{\dagger}d_{i}=\sum_{\sigma\sigma'}f_{i,\sigma}^{\dagger}\tau_{\mu,\sigma\sigma'}f_{i,\sigma'},\hspace{0.2cm}\mu=0,x,y,z,
\end{equation}
which, projected onto the charge-density and spin-density components, read
\begin{align}
\label{eq:A8}
&s_{i,0}^{\dagger}s_{i,0}+\bm{s}_{i}\hspace{0.02cm}^{\dagger}\circ\bm{s}_{i}+2d_{i}^{\dagger}d_{i}=\sum_{\sigma}f_{i,\sigma}^{\dagger}f_{i,\sigma},\\
\label{eq:A9}
&s_{i,0}^{\dagger}\bm{s}_{i}+\bm{s}_{i}\hspace{0.02cm}^{\dagger}s_{i,0}+i\bm{s}_{i}\hspace{0.02cm}^{\dagger}\bm{\times}\bm{s}_{i}=\sum_{\sigma\sigma'}f_{i,\sigma}^{\dagger}\bm{\tau}_{\sigma\sigma'}f_{i,\sigma'}.
\end{align}
These five constraints can be imposed explicitly via Lagrange multiplier terms added to the Hamiltonian~\eqref{eq:Hubbard}. Following Ref.~\onlinecite{Kotliar-PRL1986}, the Hubbard Hamiltonian~\eqref{eq:Hubbard} can be recast as
\begin{align}
\label{eq:Hubbard_SB}
H^{\textrm{SRI SB}}&=\sum_{\langle i,j\rangle,\sigma,\sigma_{1},\sigma_{2}}t_{ij}(f_{i,\sigma_{1}}^{\dagger}z_{i,\sigma_{1}\sigma}^{\dagger})(z_{j,\sigma\sigma_{2}}f_{j,\sigma_{2}})\\
&\hspace{0.5cm}+U\sum_{i}d_{i}^{\dagger}d_{i},\nonumber
\end{align}
in the Hilbert subspace of (meaningful) physical states [defined by Eqs.~\eqref{eq:A6} and~\eqref{eq:A7}]. The projector operators $\underline{z_{i}}$ are defined by Eq.~\eqref{eq:matrix_z} and describe the SB (back)flow occurring when electrons  hope between lattice sites $i\rightarrow j$: single and double occupied states contribute with two (independent) channels to the hopping of slave bosons. In the former case, the bosonic state must change from $s_{i,\sigma}^{\dagger}|0\rangle$ to $e_{i}^{\dagger}|0\rangle$, whereas in the latter the transition occurs from $d_{i}^{\dagger}|0\rangle$ to $s_{i,-\sigma}^{\dagger}|0\rangle$. Therefore, $\underline{z_{i}}\propto e_{i}^{\dagger}\underline{S_{i}}+\underline{S_{T,i}}^{\dagger}d_{i}$.\cite{Li-PRB1989} The normalization factor guarantees the conservation of the transition probability even at the mean-field level, where the aforementioned constraints are imposed only on average. Finally, within the path-integral formulation of the partition function, the SB field operators are replaced by $c$-numbers to be determined by the saddle-point solution of the functional integral. Furthermore, in the adiabatic limit for spin dynamics, spin-one Bose fields must exhibit the same spatial dependence as that of the underlying magnetization texture \cite{Fresard-EPL1991,Fresard-JCM1992}; on the contrary, spin-zero Bose fields are taken to be spatially uniform at the mean-field level.


\begin{thebibliography}{99}
\bibitem{Footnote1} In particular, spin dynamics of frustrated magnets are free from anomalies rooted in the kinetic Wess-Zumino term ascribed to ferromagnetism.
\bibitem{Footnote2} We note that the support of the weight function characterizes the mesoscopic volume over which the (volume) average is performed.
\bibitem{Footnote3} Here, \textit{optimal} means that $\{R\bm{S}_{1i}\}_{i}$ minimizes the distance with respect to the spin configuration $\{\bm{S}_{2i}\}_{i}$. We also note that the sign of $R$ (i.e., where it is proper or improper) is determined by that of the singular value $d_{3}$, see Ref.~\onlinecite{Henley-AnnPhys1984} for further details.
\bibitem{Footnote4} We note here the analogy with the $XY$ model for two-dimensional magnets, where singularities in the order parameter correspond to the emergence of magnetic vortices. The presence of these topological defects translates into the angle field being multivalued across the branch cuts connecting any two vortices of the system.
\bibitem{Footnote6} $\Phi_{m}$ vanishes in the presence of a uniform and static magnetic background. This is due to the fact that the static condition translates into $\bm{m}\propto\bm{\Omega}_{t}\equiv\bm{0}$ and uniformity means that the spatial derivatives of the SO(3) order parameter vanish, so that $\vec{\bm{\Omega}}$ is also identically zero according to Eq.~\eqref{YM_fields}. Similarly, the other terms of the Lagrangian~\eqref{eq:Lag} also vanish except for the kinetic term of the Hamiltonian, leading to a simple quadratic band dispersion for the itinerant carriers.
\bibitem{Footnote7} We note that, in Ref.~\onlinecite{Dombre-PRB1989}, the summation is also averaged by the number of sublattices since the energy density \emph{per site} is considered.
\end{thebibliography}
\end{document}